# Reconstruction of fallout deposition from U.S. atmospheric nuclear tests conducted in New Mexico and Nevada

Sébastien Philippe[1,2,*], Susan Alzner[3], Gilbert P. Compo[4,5], Mason Grimshaw[6], Megan Smith[3]

[1] Department of Nuclear Engineering and Engineering Physics, University of Wisconsin, Madison, Wisconsin

[2] Program on Science and Global Security, Princeton University, Princeton, New Jersey

[3] shift7, Washington, District of Columbia

[4] Cooperative Institute for Research in Environmental Sciences, University of Colorado Boulder, Boulder, Colorado

[5] Physical Sciences Laboratory, NOAA, Boulder, Colorado

[6] ode Partners, Rapid City, South Dakota

* Correspondence to: sphilippe@wisc.edu

## Abstract

Between 1945 and 1962, the United States conducted 101 atmospheric nuclear weapon tests in New Mexico and Nevada, dispersing radioactive fallout across the contiguous United States. We reconstruct the first 10 days of fission-product deposition from the 94 non-zero-yield tests using U.S. government test data, benchmarked particle-size distributions, and the HYSPLIT atmospheric transport model driven by high-resolution ERA5 reanalysis. Comparisons with county-level iodine-131 deposition estimates from the National Cancer Institute show good agreement for tower and surface tests and motivate refinement of the particle-size distribution parameters associated with airbursts. We aggregate per-test reconstructions into a cumulative deposition map that includes long-range fallout from the 1945 Trinity test. Trinity fallout alone reached 46 states within 10 days, consistent with contamination observed in Indiana and Iowa soon after the test. An ensemble analysis bounds meteorological uncertainty for Trinity, showing robust large-scale transport patterns despite relatively sparse 1945 observations.

## MAIN

Atmospheric nuclear weapons testing in the mid-20th century released large quantities of radioactive material to the environment[1]. In the United States, 101 atmospheric nuclear tests were conducted between 1945 and 1962, beginning with Trinity in New Mexico and followed by 100 detonations at the Nevada Test Site[2]. Seven of the tests at the Nevada Test Site led to zero or negligible fission yield and are not reconstructed in this study. Fallout from the 94 non-zero-yield atmospheric tests contaminated broad areas of the contiguous United States (CONUS) and exposed the population to ionizing radiation[3,4,5,6]. Quantifying radionuclide deposition at local, regional, and global scales remains essential for evaluating long-term environmental and health consequences and for anticipating the effects of any future nuclear weapon use[1,7,8,9].

Efforts to reconstruct fallout from U.S. tests have been undertaken for decades, particularly for Nevada detonations. Between 1975 and 2015, the Department of Energy (DOE) published reports documenting test characteristics, local fallout maps, and radionuclide compositions based on measurements and modeling[2,3,10]. In 1990, DOE compiled sparse nationwide monitoring data[4]. The National Cancer Institute (NCI) later produced county-level estimates of iodine-131 deposition and thyroid doses, while the Department of Health and Human Services (DHHS), in collaboration with the Centers for Disease Control and Prevention (CDC) and NCI, extended interpolated maps to additional radionuclides[5,11,12,13]. More recently, NCI examined Trinity fallout and cancer risks in New Mexico[14]. These studies provided valuable insights for public health but had to rely on sparse historical measurements and interpolation.

Physics-based atmospheric transport and dispersion models provide a complementary approach. By representing the transport and deposition of particles in space and time, they can extend reconstructions beyond monitoring networks and quantify deposition at higher temporal resolution. Early applications in 1990 reconstructed fallout for 13 Nevada tests using sparse observations and variational wind reconstructions on a 32-km grid, extending 18 hours and 1,200 kilometers from the test site[15]. The study also extended fallout patterns from the Trinity test in New Mexico to Colorado. Later, NCI conducted case studies of three nuclear tests[16], using the National Oceanic and Atmospheric Administration 's Hybrid Single-Particle Lagrangian Integrated Trajectory (HYSPLIT) model[17], driven by global reanalyzed meteorological fields from the NCEP-NCAR Reanalysis, which has a spatial resolution of 2.5-degree (~275 km) and temporal resolution of 6 hours[18]. HYSPLIT has been widely validated and used for applications such as tracking volcanic ash[19], wildfire smoke[20], and radionuclide releases from nuclear accidents[21]. In 2014, NOAA used HYSPLIT to reconstruct fallout from six Nevada nuclear tests using both the global NCEP–NCAR fields and a higher-resolution 12-km regional dataset created by merging global reanalysis with local western U.S. observations[22]. The enhanced resolution improved the agreement between modeled and observed local fallout. The method was later applied to the reconstruction of French nuclear tests in the Pacific, which were found to be in good agreement with declassified government data[23]. These efforts highlighted both the value of transport models and the benefits of finer spatial and/or temporal meteorological data for historical reconstruction accounting for both gravitational settling and wet deposition processes.

Recent advances in meteorological reanalysis make even higher-resolution reconstructions possible. The European Centre for Medium-Range Weather Forecasts (ECMWF) ERA5 reanalysis weather dataset now offers ~31 km spatial and 1-hour temporal resolution with 137 vertical levels extending to ~80 km (0.01 hPa) above the Earth's surface[24,25]. In 2023, ERA5 was extended back to 1940, making it the only gridded weather data utilizing direct observations of the atmospheric winds crucial to fallout dispersion that spans the complete period of U.S. atmospheric nuclear weapon testing[26,27]. It incorporates newly digitized meteorological wind observations not previously available to other efforts[28]. To help characterize meteorological uncertainty, ERA5 also provides 10 ensemble members at 0.5° resolution and 3-hour intervals. These ensembles differ from the higher-resolution ERA5 fields through perturbed boundary conditions (e.g., sea surface temperatures), perturbed observations consistent with their expected errors, and stochastically varied model parameterizations. Together, they provide a representation of random uncertainty in the otherwise unperturbed ERA5 fields[24,29].

Here, we combine ERA5 with HYSPLIT to reconstruct the first 10 days of fission-product deposition across the entire contiguous United States at a 0.1° (~10 km) resolution for the 94 non-zero-yield atmospheric tests conducted in New Mexico (Trinity) and Nevada (93 non-zero-yield tests). Our approach leverages publicly available U.S. government data about the location, time, and yield of atmospheric tests, as well as data about the heights of stabilized radioactive debris clouds and particle-size distributions. We benchmark our reconstructions against county-level iodine-131 deposition estimates obtained from the NCI, refining particle-size parameters for airbursts. We also present an ensemble analysis of Trinity using all 10 ERA5 members to bound meteorological uncertainty for our analysis. Together, these results show that coupling high-resolution reanalysis with physics-based transport modeling enables continental-scale fallout reconstructions where monitoring data are limited or unavailable and provides a framework that can be extended to other historical nuclear weapon test programs worldwide when source terms can be estimated.

## RESULTS

We reconstructed the dispersion and deposition of fission products from 94 atmospheric U.S. nuclear weapon tests across the contiguous United States with yields ranging from 200 kg to 74 kt of TNT equivalent (on kiloton of TNT is equal to $4.184\times10^{12}$ J). The first U.S. test, Trinity, was a plutonium-based device detonated on a ~30-meter-tall metallic tower in New Mexico on July 16, 1945. The other 93, conducted in Nevada between 1951 and 1962, included 48 ground or tower tests and 45 airbursts. Seven additional atmospheric tests were excluded from reconstruction: six "safety tests" with zero yield that resulted in local plutonium deposition only, and one very low-yield experiment (~36 g TNT equivalent; 1 g = $10^{-9}$ kt) for which no radioactive debris cloud was reported. Figure 1 summarizes the location and cumulative yield of the atmospheric test included in this study.

For each included test, we modeled fallout transport and reconstructed deposition over the first 10 days following detonation using NOAA's HYSPLIT model [17,22], driven by meteorological fields drawn from the ERA5 reanalysis project (1940 – present, March 2023 update)[25]. Reconstructions were initialized from a stabilized radioactive debris cloud using publicly available test parameters–location, time, yield, and cloud geometry–obtained from U.S. government reports (Supplementary Table 1) [2,3], and include particle-size-dependent transport, gravitational settling, and wet deposition. Standard particle-size distributions for ground bursts and airbursts were first obtained from the U.S. Department of Energy DELFIC code[30]. Deposition was computed on a 0.1° grid over CONUS and aggregated to produce both per-test and cumulative deposition estimates before being benchmarked to NCI deposition estimates. A 10-day period captures when deposition is concentrated and activity remains relatively high[5]. Full details of model initialization, source terms, particle properties, deposition parameterizations, and post-processing are provided in the Methods section.

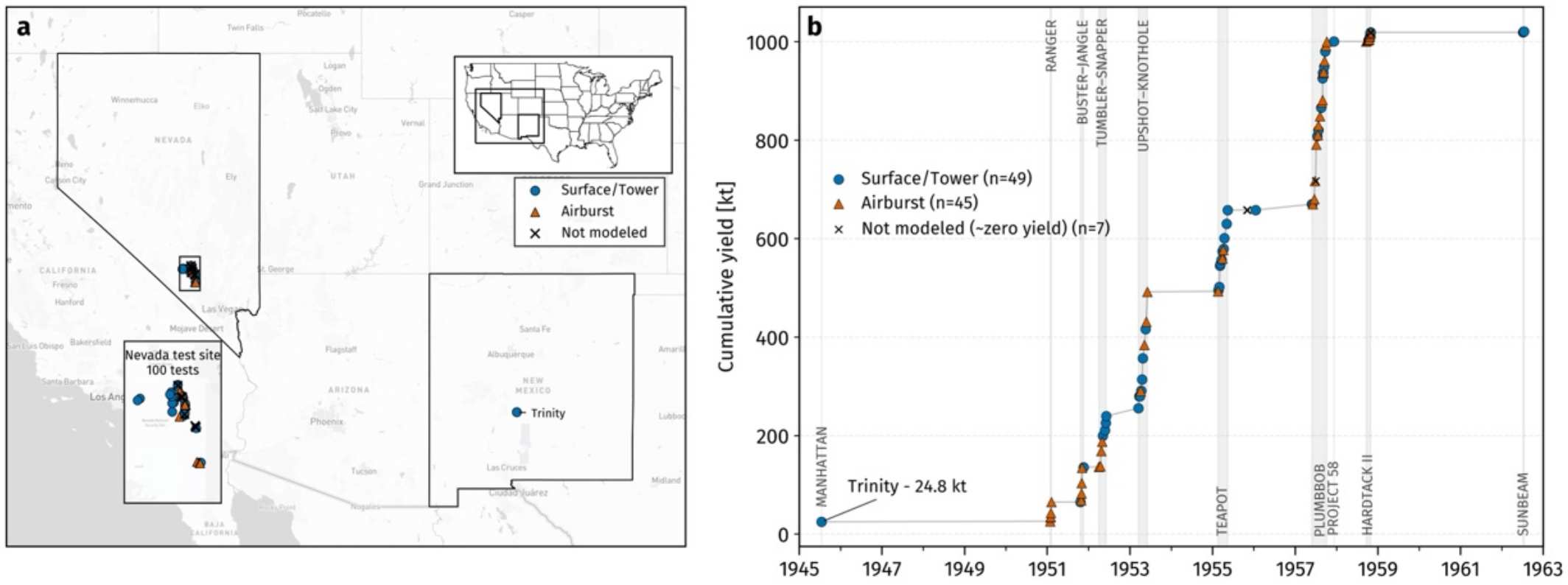

**Figure 1. Location and cumulative yield of U.S. atmospheric nuclear tests. (a)** Locations of U.S. atmospheric nuclear tests conducted between 1945 and 1962 in Nevada and New Mexico. Tests are categorized by detonation mode: surface or tower tests (blue), airbursts (orange), and safety or other tests not modeled in this study (black ×). The Nevada Test Site (inset) concentrates most tests, while the Trinity test—the first nuclear detonation—is shown in New Mexico. **(b)** Cumulative explosive yield (kilotons TNT equivalent) of U.S. atmospheric nuclear tests over time, with individual tests plotted chronologically by Operation series name and colored by detonation category. Shaded vertical bands denote major test series. Tests contributing zero or negligible yield were not modeled and are shown for completeness.

## Comparison of HYSPLIT results with historical iodine-131 deposition data

HYSPLIT deposition estimates were first compared to the National Cancer Institute (NCI) reconstruction of nationwide, county-specific iodine-131 fallout deposition resulting from atmospheric nuclear testing at the Nevada Test Site[5]. This NCI dataset provides iodine-131 deposition densities for 3,094 U.S. counties based on a subset of the atmospheric tests (66) conducted between 1951 and 1958. It was developed primarily through the reanalysis and interpolation of experimental data collected at gummed-film monitoring stations across the United States, supplemented with local deposition data and statistical modeling. These measurements involved between 40 and 95 stations operating from October 1951 to November 1958[4]

For this comparison, we used data from 51 of the 66 tests, excluding those for which deposition estimates were not derived from gummed-film data or for which fallout patterns overlapped with another test (see methods).

Consistent with the NCI methodology, we assumed a release of 6.3 $10^{15}$ Bq of iodine-131 per kiloton of fission yield (based on fast fission of plutonium-239), with the release treated as entirely particulate-bound for transport and deposition[5]. This provides a first-order approximation that is most appropriate in the near field, where iodine is predominantly particle-associated, but becomes less accurate at longer distances as elemental and organic gaseous forms contribute an increasing fraction of total iodine. A fully time-dependent treatment of iodine speciation would require a coupled chemical transport model and is beyond the scope of this study.

The NCI reconstruction adopts a similar simplifying assumption, treating iodine-131 as particle-associated for deposition calculations, reflecting the intermediate deposition behavior of particulate iodine relative to organic and elemental gaseous forms. In addition, the NCI dataset is not based on direct iodine-131 measurements but is derived from gross beta activity collected on gummed-film samplers, which primarily capture particulate fallout. The comparison presented here is therefore most consistent with a particulate-based representation of iodine transport. This approximation provides a conservative and internally consistent basis for comparison.

Figure 2 summarizes the reconstruction deposition results by county across the contiguous United States and compares HYSPLIT-based iodine-131 deposition to the NCI dataset for tower shots and airbursts.

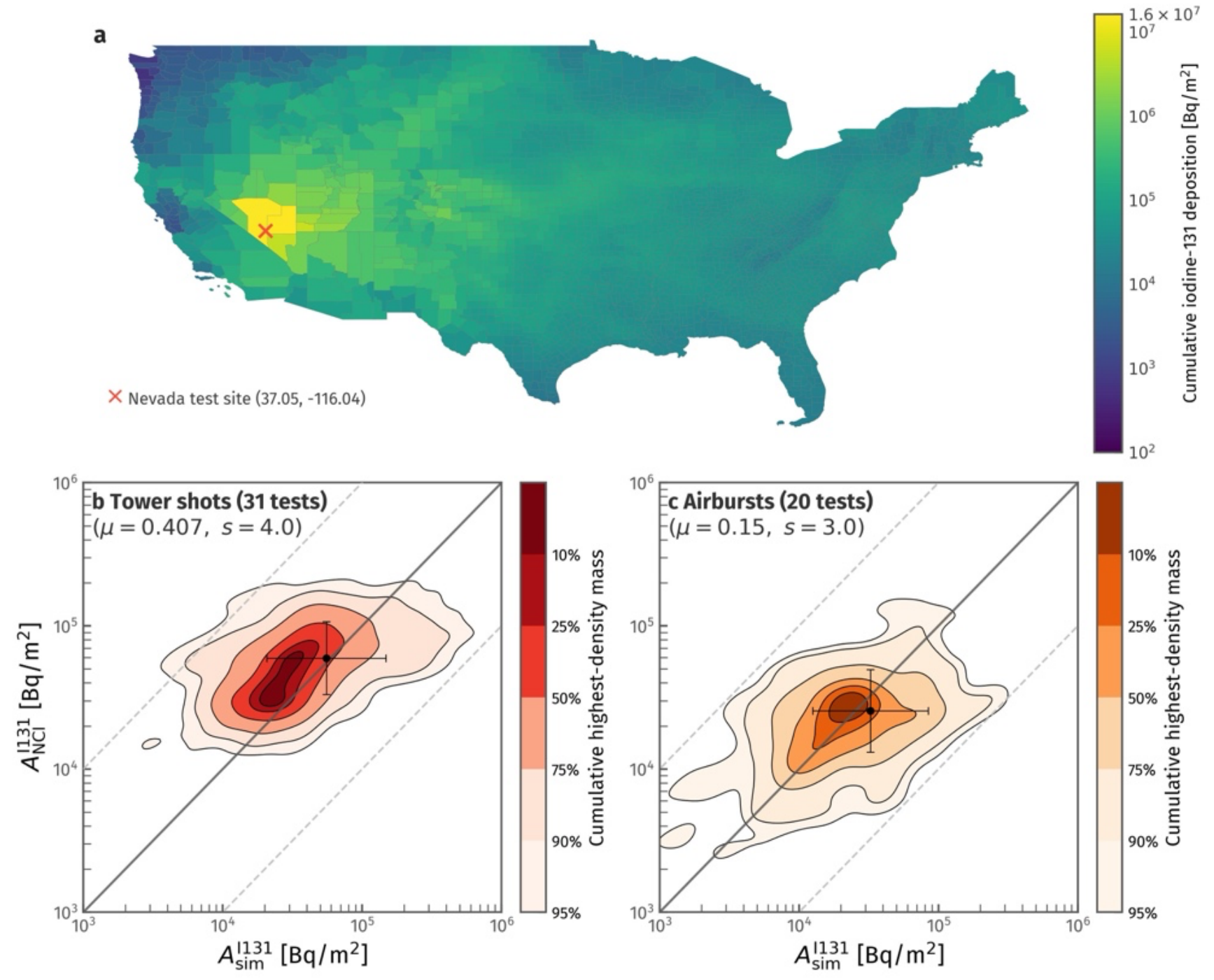

**Figure 2. Cumulative iodine-131 deposition across the United States from selected atmospheric nuclear tests. (a)** Reconstructed iodine-131 ground deposition densities (Bq m⁻²) from 51 atmospheric tests conducted at the Nevada Test Site (31 tower shots and 20 airbursts), for which National Cancer Institute (NCI) monitoring-station–based data are available. Deposition was reconstructed using HYSPLIT and aggregated to the U.S. county level. County values represent the sum of per-test mean deposition densities across the selected tests. The red cross marks the Nevada Test Site. **(b, c)** Comparison between modeled iodine-131 deposition and the NCI county-level dataset for tower shots (31 tests) and airbursts (20 tests), respectively. Each panel shows reconstructed (x-axis) versus NCI-reported (y-axis) county-level deposition densities (Bq m⁻²). The black marker indicates the mean of paired values, with asymmetric error bars representing ±1σ dispersion in log space. Shaded contours represent cumulative highest-density probability regions derived from a two-dimensional kernel density estimate computed in log-log space, illustrating where increasing fractions of counties are concentrated.

For tower shots, the reconstructed deposition agrees well with the NCI data when using the standard DELFIC particle-size distribution for ground bursts (Figure 2B). In contrast, for airbursts, the standard DELFIC distribution, defined by a median particle diameter of d=0.15 μm and geometric standard deviation s = 2, underestimates deposition by an order of magnitude.

We therefore reran the airburst reconstructions, increasing the geometric standard deviation to s = 2.5 and s = 3.0, while keeping the median diameter fixed, and compared the results to the NCI data (see Supplementary Figures 2 and 3). We find that a distribution with parameters of d=0.15 μm, s=3 yields the best agreement with the NCI measurements (Figure 2C). The larger geometric standard deviation is consistent with previous findings that the sizes of airburst particles span 3 orders of magnitude from hundredths to tens of micrometers (up to 100 μm).[31,32] We adopt these parameters for all subsequent airburst reconstructions.

Quantitative validation metrics, summarized in Table 1, support these conclusions. For tower shots, agreement between modeled and NCI iodine-131 deposition is strong across counties, with 97% of estimations within a factor of 10 of NCI values (FAC10 = 0.97) and a Pearson correlation coefficient of r = 0.50 on log10-transformed values. The mean bias is small (MBE = −0.19 in log10 space), corresponding to a modest underprediction of approximately a factor of 1.5 on average.

For airbursts, model performance is highly sensitive to the assumed particle-size distribution. Using a broader lognormal particle-size distribution (s = 3.0) yields excellent agreement (FAC10 = 0.99; r = 0.52), with negligible bias (MBE ≈ 0) and a mean absolute error corresponding to agreement within approximately a factor of 2. In contrast, the DELFIC airburst distribution leads to systematic underprediction and substantially reduced agreement (e.g., FAC10 = 0.10 and MBE = −1.63 for s = 2.0). Increasing the spread of the distribution enhances the contribution of larger particles, leading to increased deposition and improved agreement with NCI reconstructions.

While correlation coefficients are comparable across cases (r ~ 0.5), indicating that the broad spatial structure of deposition is reasonably captured, bias and error metrics show that accurate magnitude reconstruction depends critically on particle-size assumptions. The similarity in correlation across cases further indicates that improvements in model performance arise primarily from reductions in systematic bias rather than changes in spatial pattern.

**Table 1. Quantitative comparison between modeled and NCI iodine-131 deposition at the county level.** Metrics include the fraction of predictions within a factor of 10 (FAC10), 5 (FAC5), and 2 (FAC2), the Pearson correlation coefficient computed on log10-transformed values, and measures of bias (mean bias error, MBE) and error magnitude (mean absolute error, MAE), both expressed in log10 space. Results are shown for tower shots and airbursts under different assumptions for the lognormal particle-size distribution (geometric standard deviation, s). The standard DELFIC parameter for Airbursts is s=2.0.

| Cases | FAC10 | FAC5 | FAC2 | r (log10) | MBE (log10) | MAE (log10) |
|---|---|---|---|---|---|---|
| **Tower shots** | 0.97 | 0.93 | 0.59 | 0.50 | -0.19 | 0.31 |
| **Airbursts (s=2.0)** | 0.10 | 0.04 | 0.01 | 0.49 | -1.63 | 1.63 |
| **Airbursts (s=2.5)** | 0.86 | 0.55 | 0.11 | 0.49 | -0.67 | 0.68 |
| **Airbursts (s=3.0)** | 0.99 | 0.94 | 0.64 | 0.52 | +0.01 | 0.27 |

**All fission-products fallout estimates from the Trinity test**

We next focus on the 1945 Trinity test, which has historically received far less modeling attention than the later Nevada Test Site detonations. Previous reconstructions of Trinity fallout have largely relied on sparse surface observations and low-resolution meteorological data, producing only short-range deposition maps. In contrast, our reconstruction provides a 10-day nationwide estimate of all fission products deposition following Trinity, using ERA5 reanalysis fields and the same stabilized cloud and particle distribution framework (here a tower shot) applied to later tests.

Figure 3 shows the spatial distribution and time evolution of deposition estimated for Trinity. The radioactive plume initially traveled northeast from the test site, depositing material across parts of New Mexico, the Midwest, and the Great Lakes region. Within 48 hours, fallout extended across the central United States, and by day 10, deposition had occurred in 46 states. Only Washington and Oregon show no deposition. Deposition is seen to the south and west of ground zero within the first 48 hours.

These results are consistent with the limited radiological and meteorological data available for Trinity. In particular, our reconstruction is consistent with a 1987 NOAA analysis, which used available post-detonation radiological survey data and meteorological reconstructions to model fallout for the first 24 hours.[33] Our long-range fallout results also are consistent with contemporary evidence of radioactive contamination in strawboard packaging material used for X-ray films, manufactured in Indiana and Iowa in 1945, one of the earliest documented accounts of Trinity's far-field impacts (see Supplementary Figure 4).[34]

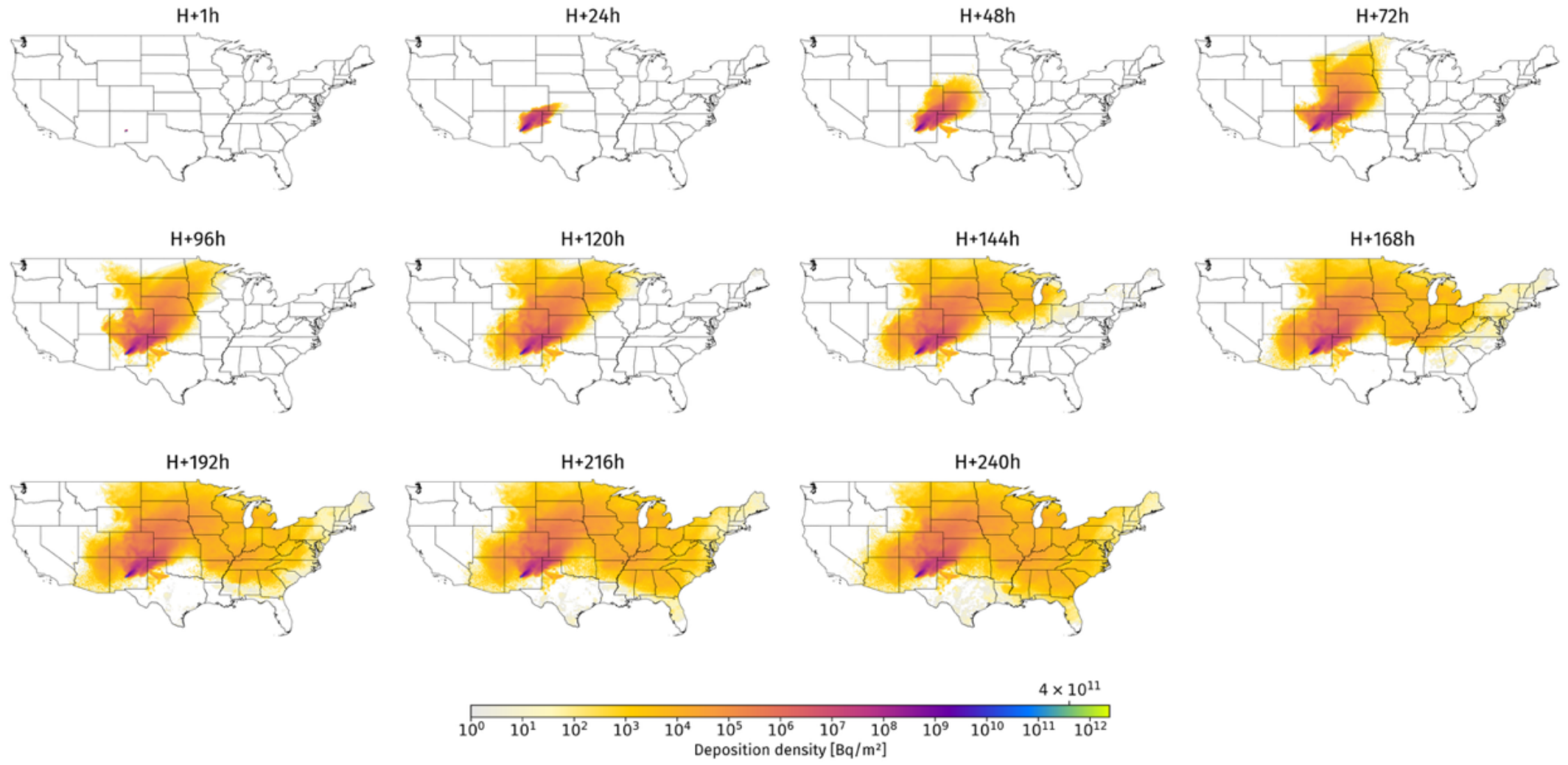

**Figure 3. Cumulative fission-product deposition density (Bq/m²) from the Trinity test over the first 10 days following detonation at 05:29 AM local time on July 16, 1945.** Maps show the spatial evolution of deposition across the contiguous United States at selected time intervals (H+1 h to H+240 h), based on reconstructions using HYSPLIT modeling driven by ERA5 meteorological reanalysis. The Trinity detonation site (33.7°N, 106.5°W; New Mexico) corresponds to the center of the highest deposition region. The fallout plume initially spread northeast from the test site, with additional deposition to the south and west within the first 48 hours. By H+240 h, cumulative deposition reached 46 states. Deposition densities are shown on a logarithmic scale to capture both local and long-range contributions (4x $10^{11}$ Bq/$m^2$ is the highest value at the Trinity Site.)

## Uncertainties and ensemble analysis

There are multiple sources of uncertainty in our deposition density estimates of the Trinity and the Nevada atmospheric tests, which can be grouped into four categories: (1) uncertainties about nuclear test parameters (e.g., fission yield, fuel type, cloud height), (2) uncertainties in the particle source term (e.g., spatial distribution, size distribution, density), (3) uncertainties related to transport modeling and post-processing (e.g., treatment of particulate versus gaseous nuclides), and (4) uncertainties in the meteorological fields. Many of these uncertainties are difficult to quantify due to limited or incomplete historical data.

In addition to meteorological variability, the reconstructed deposition fields depend on several modeling assumptions. Reported yields determine the total activity released, while cloud-height specifications set the initial vertical distribution of activity and thus influence long-range transport. Particle-size distributions are particularly important, as they control settling velocities and scavenging efficiencies. Wet deposition parameterizations introduce additional uncertainty. Although a full parametric sensitivity analysis is beyond the scope of this study, these factors define important bounds on the reconstructed deposition fields.

Regarding test parameters, we rely on public information available from U.S. government sources and the open literature. Yields remain uncertain; for example, the Trinity yield was recently reevaluated from 21 kt to 24.8kt of TNT equivalent (a ~20% increase).[35] The fissile composition of most devices is also not publicly documented, and we assume plutonium-239 for all tests. Differences in cumulative fission-product yields between plausible fissile fuels are about less than 10% for the radionuclides considered here; for example, fast fission of uranium-235 produces 2.13 $10^{21}$ Bq of iodine-131 per kt, compared with 1.96 $10^{21}$ Bq per kt for plutonium-239.[36] We also neglect contributions from neutron-induced activity in bomb debris and unfissioned fissile material, leading to conservative estimates.

Uncertainties in the source term and dry and wet deposition modeling can substantially affect predicted deposition patterns. HYSPLIT assumes constant particle sizes and densities and may underestimate settling velocities at lower altitudes while overestimating scavenging in light precipitation.[37,38] The default DELFIC parameters of the specified log-normal particle size distribution for ground and tower tests were determined from an unreferenced set of samples from Nevada Test Site fallout[30], while those for airbursts were determined from samples collected during 8 airburst tests of the Dominic (1962) test series in the north Pacific Ocean[39].

Several factors suggest that the default DELFIC airburst parameters may not be fully representative of the contiguous U.S. airburst tests analyzed here. The Dominic samples were collected exclusively from the cloud cap 2–3 hours post-detonation and from the stem 4–5 hours post-detonation, and they originate from a small number of thermonuclear tests conducted over Kiribati or the open Pacific Ocean during a single test series. As such, they may not capture the continental conditions, type of tests (fission-only as opposed to thermonuclear devices), or range of yields from Nevada Test Site airbursts , which span yields from 200 kg to 74 kt (mean 11.1 kt), compared with 200 to 10,000 kt (mean ~3,000 kt) for the Dominic samples.

Additional sources of uncertainty include imperfect theoretical assumptions, measurement limitations, and sampling constraints inherent in the original data. The samples of particle sizes utilized to determine the default DELFIC airburst parameters have a range of only about two orders of magnitude, consistent with the geometric standard deviation of 2 encoded as the default airburst parameter. This is much smaller than the 3 orders of magnitude range from hundredths to tens of micrometers (up to 100 μm) reported by others[31,32]. These broader observations support the use of a larger geometric standard deviation (GSD = 3), as adopted here, which better reflects particle-size variability not captured by the DELFIC defaults. A comprehensive exploration of this parameter space, beyond the sensitivity analyses presented in Supplementary Figures 2 and 3, is beyond the scope of the present study.

To assess meteorological uncertainty, we performed an ensemble analysis of the Trinity test using all 10 available ERA5 ensemble members. Trinity occurred in July 1945, near the beginning of the ERA5 reanalysis period, when the global meteorological observing system of largely surface observations was enhanced by the network of new and expanded upper-air observations, primarily over North America and Europe (Supplementary Figure 1). The density and quality of observations included in ERA5 increases steadily from the Trinity test across the other test periods[27]. As such, the spread of ensemble outcomes for Trinity offers a conservative upper bound on the meteorological uncertainty affecting reconstructed deposition.

Each ERA5 ensemble member represents a physically plausible realization of the atmosphere, derived from the same assimilated observations, but each observation is used with perturbed values within the data assimilation system. These perturbations are consistent with the observation errors in the high-resolution ERA5 that are assigned in the data assimilation algorithm based on decades of meteorological data assimilation and observation research. Each ensemble member also has slightly different sea surface temperature fields imposed as a boundary condition at the surface of the atmosphere. Parameters in the assimilation model itself are also stochastically perturbed to account for uncertainties in their values as specified in parameterized processes, such as convection. For each member, HYSPLIT was run with the same source terms, transport parameters, and deposition grid as the control case.

While the spatial extent and intensity of fallout varied slightly across ensemble members, all exhibited consistent large-scale patterns. Deposition reached 46 states in six members and 45 in four members. Supplementary Figure 5 shows the 10-day cumulative deposition maps for each ensemble member and the control, highlighting broad agreement in plume structure and transport direction.

Figure 4 presents a summary of reconstructed deposition densities at 85 historical fallout monitoring station locations that show non-zero values across ensemble members. These gummed-film stations, while not operational during Trinity, are representative of U.S. sites used in later testing. At most monitoring locations, ensemble deposition values cluster tightly, and the control reconstruction lies well within the interquartile range, indicating strong consistency when accounting for uncertainty in the specified meteorological fields.

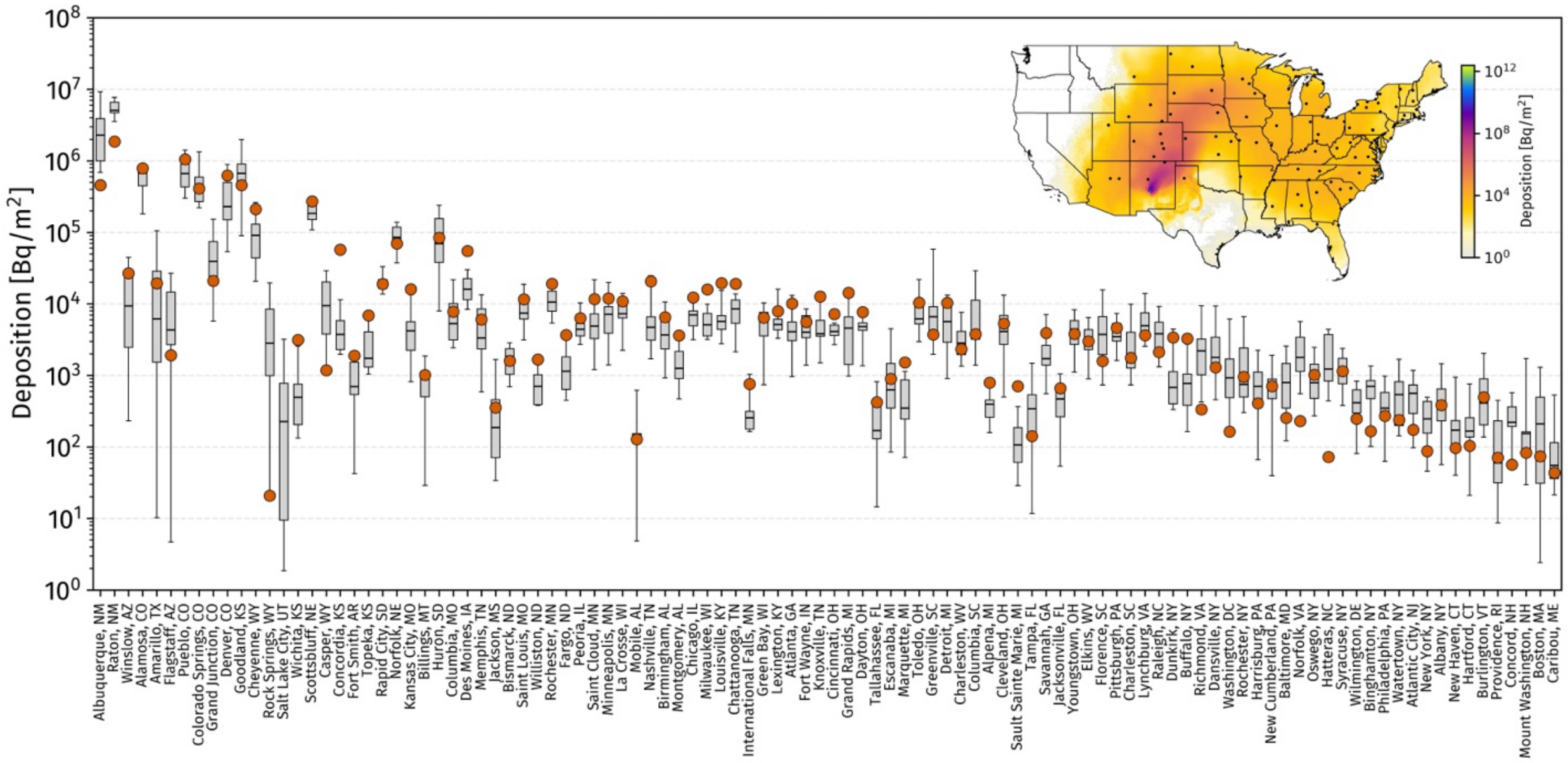

**Figure 4. Reconstructed cumulative fission-product deposition from the Trinity test across 85 historical fallout monitoring locations.** Boxplots show the distribution of 10-day deposition densities (in Bq/m²) at each station from 10 HYSPLIT reconstructions using all ERA5 ensemble members. Orange circles indicate deposition values from the control reconstruction using the main, high-resolution ERA5 reanalysis. Stations are ordered by increasing distance from Trinity ground zero. The plotted stations represent a subset of historical gummed-film monitoring sites with non-zero deposition across all ensemble members. The inset map shows the mean deposition field averaged using all 10 ensemble members. It also indicates the locations of the monitoring stations.

The largest deviation is observed at the closest station (Albuquerque, NM), which may reflect sensitivity during the early phase of fallout to the coarser 3-hour temporal and 0.5-degree spatial resolution of the ERA5 ensemble members, when particles are still highly radioactive and concentrated near the source. This highlights that near-field deposition estimates are especially sensitive to fine-scale meteorological variability in the hours immediately following detonation, variability that is not fully resolved in the lower resolution ensemble fields.

To quantify spatial agreement, we calculated pairwise Pearson correlation coefficients between deposition fields from all ensemble members, the ensemble mean, and the control. Assuming that meteorological conditions are spatially independent over 5° x 5° cells, the 20° × 60° modelling domain yields 4 x 12 = 48 degrees of freedom. This corresponds to a two-sided Pearson significance threshold of 0.36 at the $p = 0.01$ level. All correlation values exceed this threshold. Pairwise correlations between ensemble members range from 0.41 to 0.99, and from 0.41 to 0.96 between each ensemble and the control, indicating strong spatial agreement across the ensemble spread (see Supplementary Figure 6).

Together, these results demonstrate that while meteorological uncertainty introduces some variation in localized fallout estimates, the overall structure and geographic reach of the Trinity fallout plume are robust.

**Cumulative deposition density from atmospheric nuclear tests in New Mexico and Nevada**

Figure 5 shows the cumulative fission-product deposition density map across the contiguous U.S. for the first 10 days after each of the 94 non-zero yield atmospheric nuclear tests included in this study. The spatial pattern reproduces several features seen in earlier DOE and NCI deposition estimates from the 1990s, which were based on limited monitoring data and focused solely on Nevada tests. [4,18,40]

Our results newly integrate fallout from the 1945 Trinity test, highlighting its substantial contribution to national-scale deposition, particularly in New Mexico. Within 10 days, fallout from Trinity alone reached 46 states. This is consistent with its low detonation height (~30 meters) and yield of 24.8 kt, representing ~5% of the total yield from all surface and tower tests conducted in the contiguous U.S. (see Supplementary Table 1 for data on all tests).

To our knowledge, this map offers the most comprehensive, high-resolution, physics-based reconstruction to date of cumulative fallout from U.S. atmospheric nuclear tests conducted in New Mexico and Nevada.

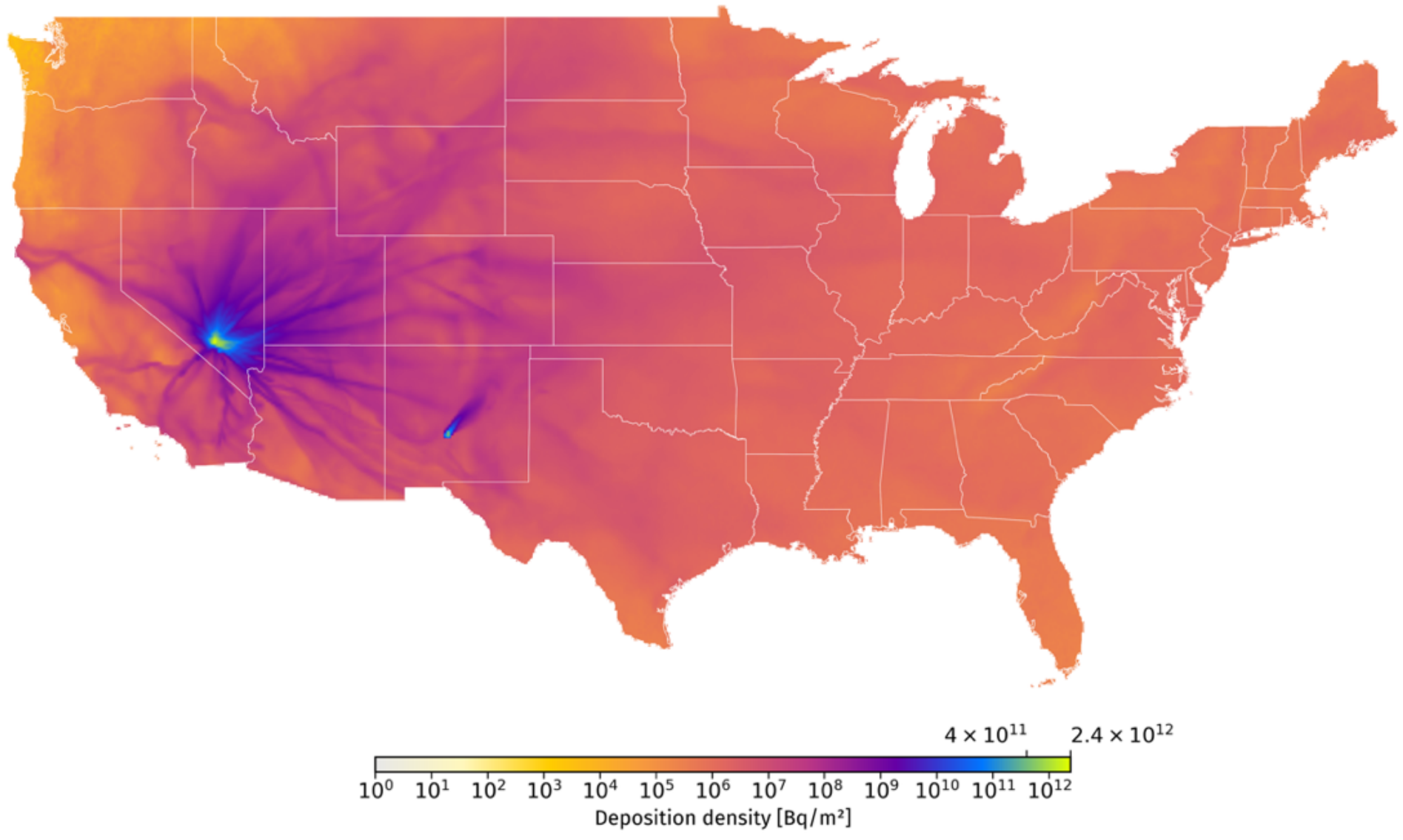

**Figure 5. Reconstructed cumulative deposition density (Bq/m$^{-2}$) of fission products from all atmospheric nuclear tests conducted in New Mexico and Nevada.** The map shows 10-day cumulative ground deposition across the contiguous United States from the 94 non-zero yield atmospheric nuclear tests included in this study. The highest deposition densities occur at the Trinity test site in New Mexico (33.7°N, 106.5°W) and the Nevada Test Site (37.1°N, 116.0°W), reflecting localized fallout from low-altitude detonations. Long-range transport of radioactive material extends across the entire contiguous U.S., consistent with patterns observed in historical monitoring campaigns. Deposition densities are shown on a logarithmic scale (2.4x$10^{12}$ Bq/m$^2$ is the highest value in the reconstruction, found at the Nevada Test Site, and 4x $10^{11}$ Bq/m$^2$ is the highest value at the Trinity Site.)

## DISCUSSION

This study presents a high-resolution, physics-based reconstruction of radioactive fallout from atmospheric nuclear weapon tests conducted in the United States between 1945 and 1962. Using modern reanalysis meteorological data (ERA5) and NOAA's HYSPLIT atmospheric transport model, we reconstruct the first 10 days of fission-product deposition for the 94 non-zero-yield tests across the contiguous United States (CONUS). The results provide an internally consistent, time-resolved estimate of both local and long-range deposition across CONUS.

We benchmarked our reconstructions against county-level iodine-131 deposition estimates from the National Cancer Institute for 51 Nevada tests. Agreement is strong for tower and surface tests using standard groundburst particle-size assumptions. For airbursts, the comparison indicates that the standard DELFIC airburst size distribution underestimates deposition and motivates a broader distribution (fixed median diameter with larger geometric spread) that yields substantially improved agreement with the NCI dataset. This benchmarking step strengthens confidence in the particle-size parameterization adopted for subsequent airburst reconstructions. Future work could perform the same benchmarking exercise with other established atmospheric transport codes to compare results across frameworks.

A central contribution of our study is the nationwide reconstruction of fallout from the Trinity test. This reconstruction shows that Trinity fallout reached 46 states within 10 days, consistent with historical observations and prior spatially resolved estimates. Ensemble-based reconstructions using all 10 ERA5 ensemble members indicate strong spatial agreement, despite limited meteorological data available for July 1945 and suggest even greater certainty for the later period tests, when more meteorological observations are incorporated into the ERA5 fields.

By aggregating deposition from all modeled tests, we have generated a new cumulative fallout map for CONUS that includes both Nevada testing and Trinity. The spatial pattern reproduces features seen in earlier DOE and NCI reconstructions based on monitoring and interpolation, while providing an independent, physics-based estimate at higher spatial and temporal resolution. Although our modeling domain was limited to the United States, the reconstructed plumes for several tests plausibly extend into Canada and Mexico, motivating future work to explicitly quantify cross-border deposition. Our approach also allows the computation and visualization of deposition for up to 213 individual radionuclides, enabling the extension of this framework to radionuclide-specific analyses and validation where suitable observational data are available.

The total deposited activity ($Bq/m^2$) reported here aggregates radionuclides with different decay timescales and therefore does not directly correspond to radiological risk. Translating radionuclide deposition into radiation dose and health outcomes require additional modeling beyond the scope of this study, including radionuclide-specific decay chains, environmental transport and resuspension, and time-dependent exposure pathways (external irradiation, inhalation, and ingestion), as well as assumptions about shielding and human behavior. The deposition fields presented here provide a physical foundation for such dose reconstruction efforts and for systematic uncertainty analyses that integrate meteorology, source-term variability, and deposition parameterizations. Further, they establish a natural bridge between physical fallout reconstruction and subsequent dose and health-impact assessment.

Taken together, our findings highlight the broad geographic extent of fallout from U.S. atmospheric nuclear testing, beginning with the very first detonation. The methodology is generalizable and can support future reconstructions of global fallout from the full set of 528 atmospheric nuclear tests conducted worldwide between 1945 and 1980, when source terms can be estimated.

## METHODS

### Atmospheric transport calculations and post-processing

We used NOAA's Hybrid Single-Particle Lagrangian Integrated Trajectory (HYSPLIT) model to reconstruct the dispersion, transport, and deposition of radioactive fallout from each atmospheric nuclear test using high-resolution ERA5 meteorological fields.

We conducted 104 HYSPLIT reconstructions in total (using the Linux version 5.2.0, January 2022): 94 for each non-zero-yield test using the high-resolution ERA5, plus 10 additional runs for Trinity using each one of the 10 ERA5 ensemble members. Each reconstruction spanned 240 hours (10 days) post detonation and used 9,000,000 particles. We adopted a 10-day integration period, consistent with the 1997 NCI iodine-131 report indicating that approximately 10 days of data were the "principal concern" to characterize fallout from an individual test[5]. A 10-day period includes concentrated deposition and relatively high activity. It balances the computational and memory requirements for the grid size and large number of particles in our reconstruction with the dominant timescales expected for atmospheric transport and particulate deposition across the contiguous United States. It is beyond the scope of this study to account for all possible fallout impacts of the U.S. atmospheric nuclear weapons tests conducted in Nevada and New Mexico.

To initialize each reconstruction, consistent with historical data, initial particle clouds were assumed to stabilize 10 minutes after detonation[3]. They were represented as a segmented vertical linear source with activity distributed among the cap, skirt, and stem of the mushroom cloud as 0.775/0.15/0.075 for surface and tower tests and 0.9712/0.0283/0.0005 for all aerial tests (e.g., balloon, airdrop, …). In general, the base of the mushroom cap is ~0.7 of the altitude of the top of the cap (see Supplementary Table 1). The base of the skirt is 1/2 and 2/3 the altitude of the base of the cap for ground/tower bursts and air bursts respectively[23].

The cloud particle sizes were first assumed to be log-normally distributed using standard parameters from the U.S. Defense Land Fallout Interpretative Code for ground and tower tests (d=0.407 µm, s=4) and for aerial tests (d=0.15 µm, s=2), where *d* is the mean particle diameter and *s* the standard deviation of the diameter's natural logarithm[30]. Particles were distributed in 100 diameter bins of equal activity (according to the 2.5 and 3rd moment respectively) and summed to a unit release. For ground and tower tests, the 2.5 moment of the distribution was used to model the first-order effect of fractionation between refractive and volatile nuclides which are typically located within the particle volume and on the particle surface, respectively[41,42]. No fractionation was assumed for aerial tests. The log-normal parameters for aerial tests were later modified after comparing the HYSPLIT results with NCI reconstruction data for iodine-131, and all tests were re-run with the final parameters d=0.15 µm and s=3 (see main text and Fig. S3).

In HYSPLIT, the particle density is kept constant over time. We assumed a particle density of 2.5 and 4.8 g/cm$^3$ for tower (soil particle substrate) and aerial tests (metallic particle substrate), respectively[23,39]. We also used the default scavenging coefficient for wet removal, applied both to below- and within-rain cloud scavenging, covering a wide range of particle sizes[43]. (In HYSPLIT, the wet removal rate constant is defined as $\beta_{wet} = 8x10^{-5}\ P^{0.79}\ s^{-1}$ with P the precipitation rate in mm/h.) For each test, we tracked hourly particle deposition up to 10 days after detonation. The results are then post-processed to obtain radionuclide deposition density[22,23].

The HYSPLIT model grid was tailored to produce fallout deposition estimates across the entire contiguous United States. It was centered at 37.0°N, 95.0°W and spanned 20° in latitude and 60° in longitude with a 0.1° × 0.1° resolution. The number of particles and grid size necessitated substantial computational and memory resources. To expedite computation, we parallelized fallout computations across 128 threads on an AMD Ryzen Threadripper PRO 5995WX (64 cores) system with 1 TB of RAM. Overall runtime totaled approximately 2 years of CPU time.

The nuclear tests in New Mexico and Nevada used a variety of fissile materials at their core (weapon grade plutonium, highly enriched uranium, uranium-233, or a combination thereof). The specific fuel for each test is not publicly available, and we assumed fission products to originate from the fast fission of plutonium for all our reconstructions (less than ~10% variation in cumulative fission yield)[36]. Local deposition density in Bq/m$^2$ was calculated by decaying the activity generated from fission for each test from the time of detonation to the time of particle arrival in each 0.1° × 0.1° grid cell, using a modified version of the HYSPLIT con2rem routine. Deposition values were then summed over all particle sizes. The con2rem source code was adapted to increase the maximum of nuclides in the source term and to account for the beginning of decay at the time of the explosion as opposed to the beginning of the modeling period. To produce the deposition density map, decay was stopped for particles at their moment of deposition on the surface.

The source term includes 213 fission products derived from England and Rider and compiled following Rolph et al.[22,38]; the full radionuclide list, half-lives, and activities per kiloton are provided in a supplementary data file. Deposition fields are computed using the complete radionuclide inventory used in the transport simulations, including noble gases. In this framework, all radionuclides are treated using a common particulate-bound transport and deposition formalism as a first-order approximation. Noble gas radionuclides (primarily xenon and krypton isotopes) account for only ~0.3% of the total initial activity at detonation, although their relative contribution increases over time as shorter-lived radionuclides decay, reaching ~10–12% over the 10-day simulation period. This simplification is retained for consistency across radionuclides and should be considered when interpreting aggregate deposited activity. The radionuclide inventory includes short-lived species (half-lives down to a few seconds), which decay predominantly prior to the onset of atmospheric transport following cloud stabilization.

**Comparison with NCI iodine-131 data**

HYSPLIT deposition estimates were compared to the National Cancer Institute (NCI) reconstruction of nationwide, county-specific iodine-131 fallout deposition resulting from atmospheric nuclear testing at the Nevada Test Site (see main text)[5]. The NCI reconstruction was developed primarily through the reanalysis and interpolation of experimental data collected at gummed-film monitoring stations across the United States, supplemented with local deposition data and statistical modeling for 66 atmospheric tests. For this comparison, we used data from 51 of the 66 tests, excluding those for which deposition estimates were not derived from gummed-film data or for which fallout patterns overlapped with another test. For tower shots, we used test numbers (21, 27, 28, 29, 30, 33, 34, 35, 37, 38, 39, 41, 51, 52, 53, 54, 55, 57, 60, 61, 62, 63, 87, 94, 96, 101, 105, 106, 110, 113, 271); for airbursts, we used test numbers (17, 18, 19, 20, 23, 24, 25, 26, 36, 40, 42, 43, 50, 90, 91, 93, 102, 111, 114, 115). Parameters for these tests are provided in Supplementary Table 1.

**Weather reanalysis data**

The ERA5 reanalysis project (1940 – present, March 2023 update) provides all the atmospheric parameters, including geopotential height, pressure, temperature, relative humidity, wind speed and direction, precipitation, and others needed to run HYSPLIT[25]. ERA5 combines historical observations into globally complete estimates of the weather state using advanced modeling and data assimilation systems. Following its March 2023 update, it is the only reanalysis data set including upper-air atmospheric observations that covers the entire 1945-1962 period. For the month of July 1945, it leverages ~2 million global observations (about 60,000 per day), split about equally between surface and upper air measurements (see Supplementary Figure 1 for a map of the observations included for July 16, 1945). The output reanalysis fields available to the public are provided on a 0.25-degree grid with a 1-hour temporal resolution on 37 pressure and one surface level. To help bound uncertainty in the meteorological inputs, we used ERA5 10 ensemble members at 0.5° resolution and 3-hour time steps for an ensemble analysis of the Trinity test[24,29].

**Acknowledgements**: The authors gratefully acknowledge the NOAA Air Resources Laboratory (ARL) for the provision of the HYSPLIT transport and dispersion model used in this publication. Hersbach, H. et al., (2023) was downloaded from the Copernicus Climate Change Service (C3S) (2023). The results contain modified Copernicus Climate Change Service information 2020. We thank H. Hersbach of the European Centre for Medium-Range Weather Forecasts (ECMWF) for providing the observations from their Observational DataBase. Neither the European Commission nor ECMWF is responsible for any use that may be made of the Copernicus information or data it contains. The authors thank S. Chanock and S. Simon of the National Cancer Institute for providing nationwide (county-specific) total iodine-131 fallout deposition data from the Nevada Test Site. Cartographic boundary files were obtained from the U.S. Census Bureau.

The authors thank Z. Statman-Weil, T. Ingold, M. Maron, and D. Hammer for database engineering, architecture and implementation, and data visualization collaborations. The authors thank R. Sayer and B. Shander for collaboration on color science and visualization. We thank F. von Hippel and Z. Mian for useful discussions and feedback on the manuscript.

S.P. acknowledges funding through grants to the Program on Science and Global Security by Andrew Carnegie Foundation, formerly the Carnegie Corporation of New York, the MacArthur Foundation, and the Ploughshares Fund. S.P., S.A., G.P.C., M.G., and M.S. were supported in part by grants from Bezos Earth Fund and the Alfred P. Sloan Foundation. G.P.C. was supported in part by NOAA cooperative agreement NA22OAR4320151 and the NOAA Physical Sciences Laboratory. The statements, findings, conclusions, and recommendations are those of the authors and do not necessarily reflect the views of NOAA or the U.S. Department of Commerce.

**Data availability**: The datasets generated during this study have been deposited in the Dryad Digital Repository (DOI: 10.5061/dryad.cvdncjtkc). Upon publication, the repository will be made publicly available. Nationwide (county-specific) total iodine-131 fallout deposition data from the Nevada Test Site are available from the U.S. National Cancer Institute. The ERA5 gridded meteorological data can be be obtained from the Copernicus Climate Change Service (C3S). Data used for Supplementary Figure 1 can be requested from the European Centre for Medium-Range Weather Forecasts.

**Code availability**: The source code for the HYSPLIT transport and dispersion model is available upon request from the NOAA Air Resources Laboratory (https://www.ready.noaa.gov/HYSPLIT_linux.php).

**Competing Interest Statement:** The authors declare no competing interests.

**Authors contributions:** SP conceptualized the research with GPC, SA and MS. SP conducted the source term evaluations, the atmospheric transport modeling, data processing, and analysis. GPC and SP conducted the meteorological data and ensemble analysis. MG and SA contributed to data visualization. SP prepared the original draft manuscript. All authors contributed to reviewing and editing the manuscript.

[14] NCI, *Study to estimate radiation doses and cancer risks resulting from exposure to radioactive fallout from the Trinity nuclear test* https://dceg.cancer.gov/research/how-we-study/exposure-assessment/trinity (2020).
[15] Cederwall, R.T., and Peterson, K.R. Meteorological modeling of arrival and deposition fallout at intermediate distances downwind of the Nevada Test Site. *Health Physics* **59**, 5, 593-601 (1990).
[16] Moroz, B.E., Beck, L., Bouville, A., and Simon, S. L. Predictions of dispersion and deposition of fallout from nuclear testing using the NOAA-HYPSLIT meteorological model. *Health Physics* **99**, 2, 252-269, (2010).
[17] Stein, A.F., Draxler, R.R, Rolph, G.D., Stunder, B.J.B., Cohen, M.D., and Ngan, F. NOAA's HYSPLIT atmospheric transport and dispersion modeling system, *Bulletin of American Meteorological Society* **96**, 2059-2077 (2015).
[18] Kalnay, E., et al. The NCEP/NCAR 40-Year Reanalysis Project. *Bulletin of American Meteorological Society* **77**, 437-472 (1996).
[19] Stunder, B. J. B., Heffter, J. L., and Draxler, R. R. Airborne volcanic ash forecast area reliability. *Weather and Forecasting* **22**, 1132-1139 (2007).
[20] Rolph, G. D. et al. Description and verification of the NOAA smoke forecasting system: the 2007 fire season. *Weather and Forecasting* **24**, 361-378 (2009).
[21] Draxler, R. R., and Rolph, G. D. Evaluation of the Transfer Coefficient Matrix (TCM) approach to model the atmospheric radionuclide air concentrations from Fukushima. *Journal of Geophysical Research, Atmospheres* **117**, D05107, 10.1029/2011JD017205 (2012).
[22] Rolph, G. D., Ngan, F., and Draxler, R. R. Modeling the fallout from stabilized nuclear clouds using the HYSPLIT atmospheric dispersion model. *Journal of environmental radioactivity* **136**, 41-55 (2014).
[23] Philippe, S., Schoenberger, S. and Ahmed, A. Radiation exposures and compensation of victims of French atmospheric nuclear tests in Polynesia. *Science and Global Security*, **30**:2, 62-94 (2022).
[24] Hersbach, H. et al. The ERA5 global reanalysis. *Quarterly Journal of the Royal Meteorological Society*, **146**(730), 1999-2049 (2020).
[25] Hersbach, H. et al. ERA5 hourly data on single levels from 1940 to present. Copernicus Climate Change Service (C3S) Climate Data Store (CDS) 10.24381/cds.adbb2d47 (2023).
[26] Hersbach, H. ERA5 reanalysis now available from 1940. *ECWMF Newsletter* **175** https://www.ecmwf.int/en/newsletter/175/news/era5-reanalysis-now-available-1940 (2023).
[27] Soci, C. et al. The ERA5 global reanalysis from 1940 to 2022. *Quarterly Journal of the Royal Meteorological Society* **150**, 4014-4048 (2024).
[28] Brönnimann, S. et al. A roadmap to climate data rescue services. *Geoscience Data Journal* **5**, 28-39, (2018).
[29] Isaksen, L. et al. Ensemble of data assimilations at ECMWF. ECMWF Tech. Memo. 636 (ECMWF, Reading, UK, 2010).
[30] Norment, H. G. DELFIC: Department of Defense Fallout Prediction System. Volume I-Fundamentals. Atmospheric Science Associates, Bedford MA, 1-101 (1979).
[31] Glasstone S. and Dolan P.J. *The Effects of Nuclear Weapons*, Third Edition, 1-662 (U.S. Department of Defense, 1977).
[32] Martin, G. Les atolls de Mururoa et Fangataufa. Les expérimentations nucléaires, aspect radiologique, CEA-R-6136, 1-761 (Commissariat à l'Energie Atomique, 2007).
[33] Quinn, V.E. *Analysis of nuclear test Trinity radiological and meteorological data*. No. NVO-313. National Oceanic and Atmospheric Administration, Las Vegas, NV (USA). Weather Service Nuclear Support Office, 1-82 (1987).
[34] Webb, J. H. The fogging of photographic film by radioactive contaminants in cardboard packaging materials. *Physical Review* **76**, no. 3, 375-380 (1949).

[35] Selby, H. D. et al. A new yield assessment for the Trinity nuclear test, 75 years later. *Nuclear Technology* **207**, no. sup1, 321-325 (2021).
[36] England, T. R., and Rider, B. F. Evaluation and compilation of fission product yields 1993. No. LA-SUB--94-170. (Los Alamos National Lab., United States, 1995).
[37] Draxler, R. et al. World Meteorological Organization's model simulations of the radionuclide dispersion and deposition from the Fukushima Daiichi nuclear power plant accident, *Journal of environmental radioactivity* **139**, 172–184 (2015).
[38] Quérel, A., Quélo, D., Roustan, Y. and Mathieu, A. Sensitivity study to select the wet deposition scheme in an operational atmospheric transport model, *Journal of Environmental Radioactivity* **237**, 106712 (2021).
[39] Nathans, M. W., Thews, R., Holland, W. D., and Benson, P. A. Particle sized in clouds from nuclear airbursts, *Journal of Geophysical Research* **75**, 7559-7572 (1970).
[40] Beck, H. L., and Bennett, B. G. Historical overview of atmospheric nuclear weapons testing and estimates of fallout in the continental United States. *Health Physics* **82**, no. 5, 591-608 (2002).
[41] Bigelow Jr, W. S. Far field fallout prediction techniques. Air Force Institute of Technology, PhD Thesis, 1-172 (1983).
[42] Conners, S. P. Aircrew dose and engine dust ingestion from nuclear cloud penetration. PhD diss., Air Force Institute of Technology, 1-156 (1986).
[43] Sportisse, B. A review of parameterizations for modelling dry deposition and scavenging of radionuclides. *Atmospheric Environment* **41**, no. 13, 2683-2698 (2007).

**Supplementary information for**

# Reconstruction of fallout deposition from U.S. atmospheric nuclear tests conducted in New Mexico and Nevada

Sébastien Philippe[1,2,*], Susan Alzner[3], Gilbert P. Compo[4,5], Mason Grimshaw[6], Megan Smith[3]

[1] Department of Nuclear Engineering and Engineering Physics, University of Wisconsin, Madison, Wisconsin

[2] Program on Science and Global Security, Princeton University, Princeton, New Jersey

[3] shift7, Washington, District of Columbia

[4] Cooperative Institute for Research in Environmental Sciences, University of Colorado Boulder, Boulder, Colorado

[5] Physical Sciences Laboratory, NOAA, Boulder, Colorado

[6] Ode Partners, Rapid City, South Dakota

* Correspondence to: sphilippe@wisc.edu

**This file includes one supplementary table and six supplementary figures.**

**Supplementary Table 1. List of 101 atmospheric nuclear tests conducted in New Mexico and Nevada with key reconstruction parameters.** Tests marked (*) were not included in this study (see main text).

| Test | U.S. # | Operation | Name | Date | Type | Time (GMT) | Latitude | Longitude | Height [m AGL] | Yield [kt] | Cloud top [m AGL] | Cloud bottom [m AGL] |
|---|---|---|---|---|---|---|---|---|---|---|---|---|
| 1 | 1 | MANHATTAN | Trinity | 7/16/45 | Tower | 11:29 | 33.6773 | -106.4754 | 30.5 | 24.8 | 9259 | 1822 |
| 2 | 7 | RANGER | Able | 1/27/51 | Airdrop | 13:45 | 36.8051 | -115.9502 | 323.1 | 1 | 4225 | 2841 |
| 3 | 8 | RANGER | Baker | 1/28/51 | Airdrop | 13:52 | 36.8051 | -115.9502 | 329.2 | 8 | 9711 | 6531 |
| 4 | 9 | RANGER | Easy | 2/1/51 | Airdrop | 13:47 | 36.8051 | -115.9502 | 329.2 | 1 | 2853 | 1919 |
| 5 | 10 | RANGER | Baker-2 | 2/2/51 | Airdrop | 13:49 | 36.8051 | -115.9502 | 335.3 | 8 | 7577 | 5096 |
| 6 | 11 | RANGER | Fox | 2/6/51 | Airdrop | 13:47 | 36.8051 | -115.9502 | 437.4 | 22 | 12149 | 7273 |
| 7 | 16 | BUSTER | Able | 10/22/51 | Tower | 14:00 | 37.0833 | -116.0247 | 30.5 | 0.1 | 1168 | 771 |
| 8 | 17 | BUSTER | Baker | 10/28/51 | Airdrop | 15:20 | 37.0850 | -116.0211 | 340.8 | 3.5 | 8384 | 5732 |
| 9 | 18 | BUSTER | Charlie | 10/30/51 | Airdrop | 15:00 | 37.0850 | -116.0213 | 345 | 14 | 11219 | 6952 |
| 10 | 19 | BUSTER | Dog | 11/1/51 | Airdrop | 15:30 | 37.0846 | -116.0209 | 431.9 | 21 | 12743 | 8171 |
| 11 | 20 | BUSTER | Easy | 11/5/51 | Airdrop | 16:30 | 37.0919 | -116.0253 | 400.5 | 31 | 13953 | 9381 |
| 12 | 21 | JANGLE | Sugar | 11/19/51 | Surface | 17:00 | 37.1315 | -116.0394 | 1.2 | 1.2 | 3287 | 2068 |
| 13 | 23 | TUMBLER-SNAPPER | Able | 4/1/52 | Airdrop | 17:00 | 36.7980 | -115.9364 | 241.7 | 1 | 4000 | 2690 |
| 14 | 24 | TUMBLER-SNAPPER | Baker | 4/15/52 | Airdrop | 17:30 | 37.0841 | -116.0203 | 338 | 1 | 3507 | 1770 |
| 15 | 25 | TUMBLER-SNAPPER | Charlie | 4/22/52 | Airdrop | 17:30 | 37.0844 | -116.0211 | 1050.7 | 31 | 11524 | 8171 |
| 16 | 26 | TUMBLER-SNAPPER | Dog | 5/1/52 | Airdrop | 16:30 | 37.0841 | -116.0211 | 317 | 19 | 12133 | 7256 |
| 17 | 27 | TUMBLER-SNAPPER | Easy | 5/7/52 | Tower | 12:15 | 37.0530 | -116.1035 | 91.4 | 12 | 9044 | 5711 |
| 18 | 28 | TUMBLER-SNAPPER | Fox | 5/25/52 | Tower | 12:00 | 37.0955 | -116.1036 | 91.4 | 11 | 11184 | 7062 |
| 19 | 29 | TUMBLER-SNAPPER | George | 6/1/52 | Tower | 11:55 | 37.0955 | -116.1036 | 91.4 | 15 | 10050 | 6347 |
| 20 | 30 | TUMBLER-SNAPPER | How | 6/5/52 | Tower | 11:55 | 37.1384 | -116.1183 | 91.4 | 14 | 11372 | 7181 |
| 21 | 33 | UPSHOT-KNOTHOLE | Annie | 3/17/53 | Tower | 13:20 | 37.0955 | -116.1036 | 91.4 | 16 | 11270 | 7307 |
| 22 | 34 | UPSHOT-KNOTHOLE | Nancy | 3/24/53 | Tower | 13:10 | 37.0955 | -116.1036 | 91.4 | 24 | 11336 | 6611 |
| 23 | 35 | UPSHOT-KNOTHOLE | Ruth | 3/31/53 | Tower | 13:00 | 37.0853 | -116.0221 | 93 | 0.2 | 2926 | 2042 |
| 24 | 36 | UPSHOT-KNOTHOLE | Dixie | 4/6/53 | Airdrop | 15:30 | 37.0846 | -116.0207 | 1835.5 | 11 | 12489 | 8832 |
| 25 | 37 | UPSHOT-KNOTHOLE | Ray | 4/11/53 | Tower | 12:45 | 37.0989 | -116.0933 | 30.5 | 0.2 | 2674 | 1120 |
| 26 | 38 | UPSHOT-KNOTHOLE | Badger | 4/18/53 | Tower | 12:35 | 37.1384 | -116.1183 | 91.4 | 23 | 9604 | 5642 |
| 27 | 39 | UPSHOT-KNOTHOLE | Simon | 4/25/53 | Tower | 12:30 | 37.0530 | -116.1035 | 91.4 | 43 | 12119 | 8157 |
| 28 | 40 | UPSHOT-KNOTHOLE | Encore | 5/8/53 | Airdrop | 15:30 | 36.7979 | -115.9298 | 738.5 | 27 | 11864 | 7597 |
| 29 | 41 | UPSHOT-KNOTHOLE | Harry | 5/19/53 | Tower | 12:05 | 37.0403 | -116.0262 | 91.4 | 32 | 11733 | 7161 |
| 30 | 42 | UPSHOT-KNOTHOLE | Grable | 5/25/53 | Airburst | 15:30 | 36.7979 | -115.9298 | 159.7 | 15 | 9730 | 6073 |
| 31 | 43 | UPSHOT-KNOTHOLE | Climax | 6/4/53 | Airdrop | 11:15 | 37.0875 | -116.0192 | 406.6 | 61 | 11788 | 9441 |
| 32 | 50 | TEAPOT | Wasp | 2/18/55 | Airdrop | 20:00 | 37.0865 | -116.0227 | 232.3 | 1 | 5275 | 3141 |
| 33 | 51 | TEAPOT | Moth | 2/22/55 | Tower | 13:45 | 37.0955 | -116.1036 | 91.4 | 2 | 6149 | 3619 |
| 34 | 52 | TEAPOT | Tesla | 3/1/55 | Tower | 13:30 | 37.1254 | -116.0484 | 91.4 | 7 | 7918 | 4352 |
| 35 | 53 | TEAPOT | Turk | 3/7/55 | Tower | 13:20 | 37.1384 | -116.1183 | 152.4 | 43 | 12256 | 9330 |
| 36 | 54 | TEAPOT | Hornet | 3/12/55 | Tower | 13:20 | 37.0403 | -116.0262 | 91.4 | 4 | 10056 | 7252 |
| 37 | 55 | TEAPOT | Bee | 3/22/55 | Tower | 13:05 | 37.0948 | -116.0246 | 152.4 | 8 | 10807 | 7698 |
| 38 | 57 | TEAPOT | Apple-1 | 3/29/55 | Tower | 12:55 | 37.0955 | -116.1036 | 152.4 | 14 | 8440 | 5575 |
| 39 | 58 | TEAPOT | Wasp Prime | 3/29/55 | Airdrop | 18:00 | 37.0865 | -116.0227 | 225.2 | 3 | 8475 | 5700 |
| 40 | 59 | TEAPOT | HA | 4/6/55 | Airdrop | 18:00 | 37.0287 | -116.0587 | 9931.1 | 3 | 15533 | 10446 |
| 41 | 60 | TEAPOT | Post | 4/9/55 | Tower | 12:30 | 37.1221 | -116.0353 | 91.4 | 2 | 3433 | 2366 |
| 42 | 61 | TEAPOT | MET | 4/15/55 | Tower | 19:15 | 36.7979 | -115.9298 | 121.9 | 22 | 11345 | 8755 |
| 43 | 62 | TEAPOT | Apple-2 | 5/5/55 | Tower | 12:10 | 37.0530 | -116.1035 | 152.4 | 29 | 14254 | 9225 |
| 44 | 63 | TEAPOT | Zucchini | 5/15/55 | Tower | 12:00 | 37.0948 | -116.0246 | 152.4 | 28 | 10898 | 6387 |
| 45 | 64* | Project 56 | No. 1 | 11/1/55 | Surface | 22:10 | 36.9844 | -115.9616 | 0.9 | 0 | n/a | n/a |
| 46 | 65* | Project 56 | No. 2 | 11/3/55 | Surface | 21:15 | 36.9801 | -115.9595 | 0.9 | 0 | n/a | n/a |
| 47 | 66* | Project 56 | No. 3 | 11/5/55 | Surface | 19:55 | 36.9757 | -115.9575 | 0.9 | $3.6\ 10^{-8}$ | n/a | n/a |
| 48 | 67 | Project 56 | No. 4 | 1/18/56 | Surface | 21:30 | 36.9713 | -115.9554 | 0.9 | 0.001 | 914 | 610 |
| 49 | 87 | PLUMBBOB | Boltzmann | 5/28/57 | Tower | 11:55 | 37.0948 | -116.0245 | 152.4 | 12 | 8768 | 5720 |
| 50 | 88 | PLUMBBOB | Franklin | 6/2/57 | Tower | 11:55 | 37.0955 | -116.1036 | 91.4 | 0.14 | 3863 | 3040 |
| 51 | 89 | PLUMBBOB | Lassen | 6/5/57 | Balloon | 11:45 | 37.1347 | -116.0416 | 152.4 | 0.0005 | 722 | 486 |
| 52 | 90 | PLUMBBOB | Wilson | 6/18/57 | Balloon | 11:45 | 37.1347 | -116.0416 | 152.4 | 10 | 9379 | 6331 |
| 53 | 91 | PLUMBBOB | Priscilla | 6/24/57 | Balloon | 13:30 | 36.7979 | -115.9298 | 213.4 | 37 | 12169 | 6378 |
| 54 | 92* | PLUMBBOB | Coulomb-A | 7/1/57 | Surface | 17:30 | 37.0412 | -116.0269 | 0 | 0 | n/a | n/a |
| 55 | 93 | PLUMBBOB | Hood | 7/5/57 | Balloon | 11:40 | 37.1347 | -116.0416 | 457.2 | 74 | 13341 | 9379 |
| 56 | 94 | PLUMBBOB | Diablo | 7/15/57 | Tower | 11:30 | 37.1502 | -116.1095 | 152.4 | 17 | 8392 | 4734 |
| 57 | 95 | PLUMBBOB | John | 7/19/57 | Rocket | 14:00 | 37.1604 | -116.0538 | 4520.2 | 2 | 12104 | 7644 |
| 58 | 96 | PLUMBBOB | Kepler | 7/24/57 | Tower | 11:50 | 37.0955 | -116.1036 | 152.4 | 10 | 7221 | 4783 |
| 59 | 97 | PLUMBBOB | Owens | 7/25/57 | Balloon | 13:30 | 37.1347 | -116.0416 | 152.4 | 9.7 | 9383 | 4811 |
| 60 | 99 | PLUMBBOB | Stokes | 8/7/57 | Balloon | 12:25 | 37.0865 | -116.0248 | 457.2 | 19 | 10002 | 6954 |
| 61 | 101 | PLUMBBOB | Shasta | 8/18/57 | Tower | 12:00 | 37.1279 | -116.1073 | 152.4 | 17 | 8417 | 3540 |
| 62 | 102 | PLUMBBOB | Doppler | 8/23/57 | Balloon | 12:00 | 37.0865 | -116.0248 | 457.2 | 11 | 10293 | 5721 |
| 63 | 104 | PLUMBBOB | Franklin Prime | 8/30/57 | Balloon | 12:40 | 37.0865 | -116.0248 | 228.6 | 4.7 | 8478 | 5125 |
| 64 | 105 | PLUMBBOB | Smoky | 8/31/57 | Tower | 12:30 | 37.1872 | -116.0687 | 213.4 | 44 | 10217 | 6452 |
| 65 | 106 | PLUMBBOB | Galileo | 9/2/57 | Tower | 12:40 | 37.0530 | -116.1035 | 152.4 | 11 | 9982 | 3886 |
| 66 | 107 | PLUMBBOB | Wheeler | 9/6/57 | Balloon | 12:45 | 37.1347 | -116.0416 | 152.4 | 0.197 | 3892 | 2978 |
| 67 | 108 | PLUMBBOB | Coulomb-B | 9/6/57 | Surface | 20:05 | 37.0429 | -116.0271 | 0.9 | 0.3 | 4257 | 2688 |
| 68 | 109 | PLUMBBOB | Laplace | 9/8/57 | Balloon | 13:00 | 37.0865 | -116.0248 | 228.6 | 1 | 4820 | 2991 |
| 69 | 110 | PLUMBBOB | Fizeau | 9/14/57 | Tower | 16:45 | 37.0334 | -116.0322 | 152.4 | 11 | 10964 | 7001 |
| 70 | 111 | PLUMBBOB | Newton | 9/16/57 | Balloon | 12:50 | 37.0865 | -116.0248 | 457.2 | 12 | 8478 | 4515 |
| 71 | 113 | PLUMBBOB | Whitney | 9/23/57 | Tower | 12:30 | 37.1384 | -116.1183 | 152.4 | 19 | 7777 | 4119 |
| 72 | 114 | PLUMBBOB | Charleston | 9/28/57 | Balloon | 13:00 | 37.1347 | -116.0416 | 457.2 | 12 | 8469 | 4811 |
| 73 | 115 | PLUMBBOB | Morgan | 10/7/57 | Balloon | 13:00 | 37.1347 | -116.0416 | 152.4 | 8 | 10908 | 6640 |
| 74 | 117 | PROJECT58 | Coulomb-C | 12/9/57 | Surface | 20:00 | 37.0482 | -116.0252 | 0 | 0.5 | 2728 | 1723 |
| 75 | 160 | HARDTACKII | Eddy | 9/19/58 | Balloon | 14:00 | 37.0865 | -116.0248 | 152.4 | 0.083 | 2077 | 1010 |
| 76 | 165 | HARDTACKII | Mora | 9/29/58 | Balloon | 14:05 | 37.0865 | -116.0248 | 457.2 | 2 | 4363 | 1772 |
| 77 | 166 | HARDTACKII | Hidalgo | 10/5/58 | Balloon | 14:10 | 37.0865 | -116.0248 | 114.9 | 0.077 | 2382 | 1163 |
| 78 | 169 | HARDTACKII | Quay | 10/10/58 | Tower | 14:30 | 37.0948 | -116.0245 | 30.5 | 0.079 | 1754 | 992 |
| 79 | 170 | HARDTACKII | Lea | 10/13/58 | Balloon | 13:20 | 37.0865 | -116.0248 | 457.2 | 1.4 | 3906 | 2382 |
| 80 | 172 | HARDTACKII | Hamilton | 10/15/58 | Tower | 16:00 | 36.8023 | -115.9331 | 15.2 | 0.0012 | 890 | 433 |
| 81 | 174 | HARDTACKII | Dona Ana | 10/16/58 | Balloon | 14:20 | 37.0865 | -116.0248 | 137.2 | 0.037 | 2077 | 705 |
| 82 | 175 | HARDTACKII | Vesta | 10/17/58 | Surface | 23:00 | 37.1226 | -116.0354 | 0 | 0.024 | 1760 | 1111 |
| 83 | 176 | HARDTACKII | Rio Arriba | 10/18/58 | Tower | 14:25 | 37.0409 | -116.0268 | 22.3 | 0.09 | 2893 | 2131 |
| 84 | 178 | HARDTACKII | Socorro | 10/22/58 | Balloon | 13:30 | 37.0865 | -116.0248 | 442 | 6 | 6649 | 4820 |
| 85 | 179 | HARDTACKII | Wrangell | 10/22/58 | Balloon | 16:50 | 36.7979 | -115.9298 | 457.2 | 0.115 | 2110 | 1196 |
| 86 | 180* | HARDTACKII | Oberon | 7/1/57 | Tower | 20:30 | 37.1828 | -116.0693 | 7.6 | 0 | n/a | n/a |
| 87 | 181 | HARDTACKII | Rushmore | 10/22/58 | Balloon | 23:40 | 37.1347 | -116.0416 | 152.4 | 0.188 | 457 | 307 |
| 88 | 182 | HARDTACKII | Catron | 10/24/58 | Tower | 15:00 | 37.0428 | -116.0278 | 22.3 | 0.021 | 1364 | 297 |
| 89 | 183 | HARDTACKII | Juno | 10/24/58 | Surface | 16:01 | 37.1233 | -116.0387 | 0 | 0.0017 | 393 | 248 |
| 90 | 184 | HARDTACKII | Ceres | 10/26/58 | Tower | 4:00 | 37.1813 | -116.0695 | 7.6 | 0.0007 | 479 | 303 |
| 91 | 185 | HARDTACKII | Sanford | 10/26/58 | Balloon | 10:20 | 36.7979 | -115.9298 | 457.2 | 4.9 | 6987 | 2872 |
| 92 | 186 | HARDTACKII | De Baca | 10/26/58 | Balloon | 16:00 | 37.0865 | -116.0248 | 457.2 | 2.2 | 4058 | 1772 |
| 93 | 187 | HARDTACKII | Chaves (Chavez) | 10/27/58 | Tower | 14:30 | 37.0446 | -116.0307 | 15.8 | 0.0006 | 754 | 476 |
| 94 | 189* | HARDTACKII | Mazama | 10/30/58 | Tower | 11:20 | 37.1265 | -116.0419 | 15.2 | 0 | n/a | n/a |
| 95 | 190 | HARDTACKII | Humboldt | 10/29/58 | Tower | 14:45 | 37.0476 | -116.0254 | 7.6 | 0.0078 | 1058 | 601 |
| 96 | 191 | HARDTACKII | Santa Fe | 10/30/58 | Balloon | 3:00 | 37.0865 | -116.0248 | 457.2 | 1.3 | 4211 | 2687 |
| 97 | 192* | HARDTACKII | Ganymede | 10/30/58 | Surface | 11:00 | 37.1265 | -116.0479 | 0 | 0 | n/a | n/a |
| 98 | 194 | HARDTACKII | Titania | 10/30/58 | Tower | 20:34 | 37.1772 | -116.0701 | 7.6 | 0.0002 | 487 | 307 |
| 99 | 265 | SUNBEAM | Little Feller II | 7/7/62 | Surface | 17:00 | 37.1191 | -116.3037 | 0.9 | 0.022 | 1790 | 1130 |
| 100 | 271 | SUNBEAM | Small Boy | 7/14/62 | Tower | 18:30 | 36.8025 | -115.9259 | 3 | 1.65 | 4853 | 3065 |
| 101 | 272 | SUNBEAM | Little Feller I | 7/17/62 | Surface | 17:00 | 37.1096 | -116.3182 | 0.9 | 0.018 | 1770 | 1118 |

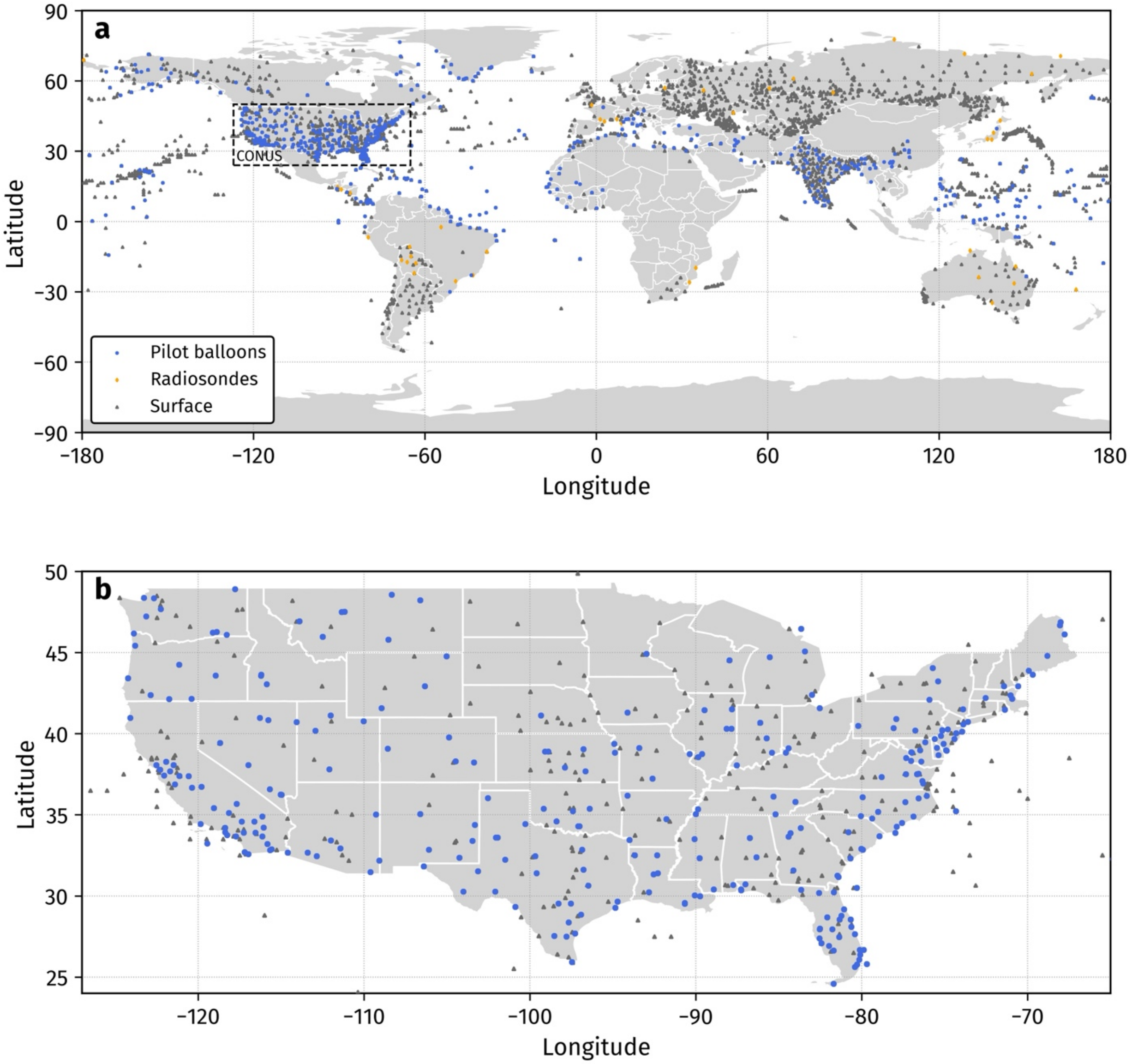

**Supplementary Figure 1. ERA5 surface and upper air (pilot balloons and radiosondes) observations for July 16, 1945.** (**a**) World-wide observations. (**b**) Details for the region of interest in this study.

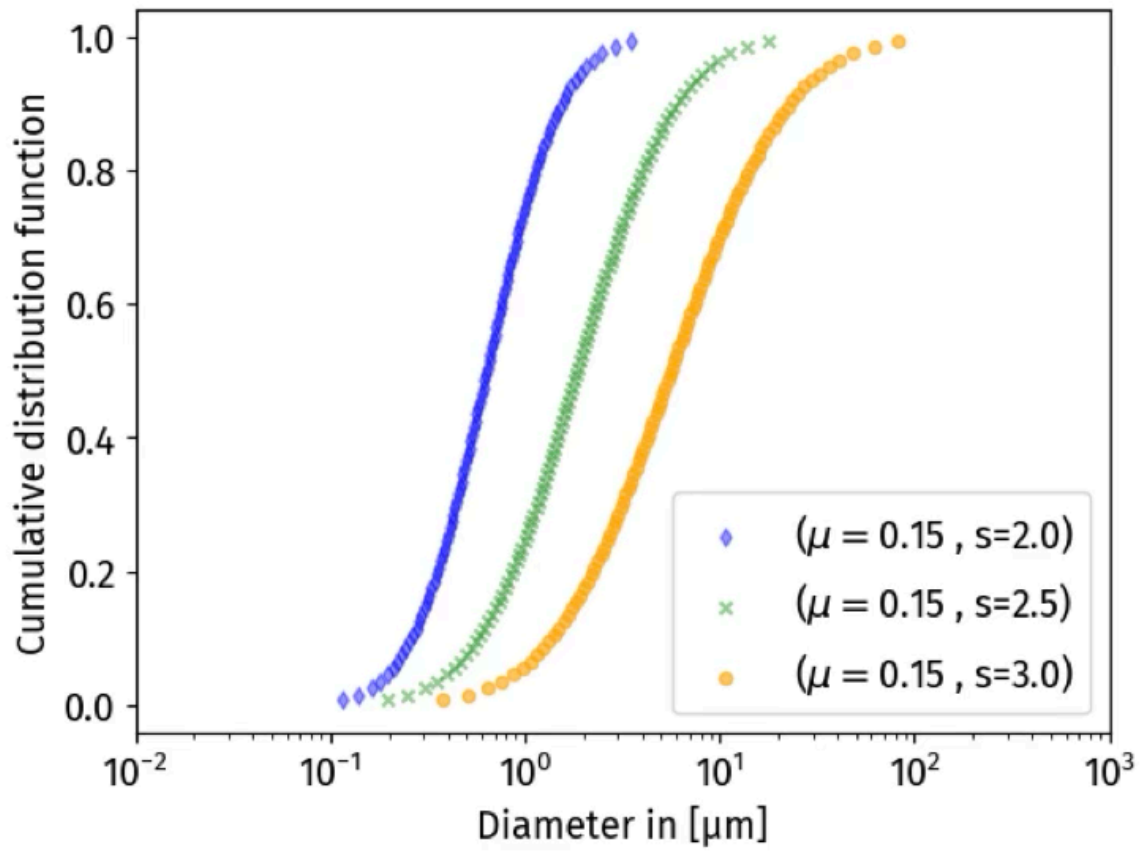

**Supplementary Figure 2. Cumulative distribution function (CDF) of particle diameters for three log-normal size distributions used in airburst reconstructions.** All distributions share the same median particle diameter (μ = 0.15 μm) but differ in geometric standard deviation (s = 2.0, 2.5, and 3.0). Increasing the standard deviation parameter (s) results in a broader distribution and a greater fraction of larger particles. These distributions were tested to benchmark model output against National Cancer Institute iodine-131 data (see Figure 1).

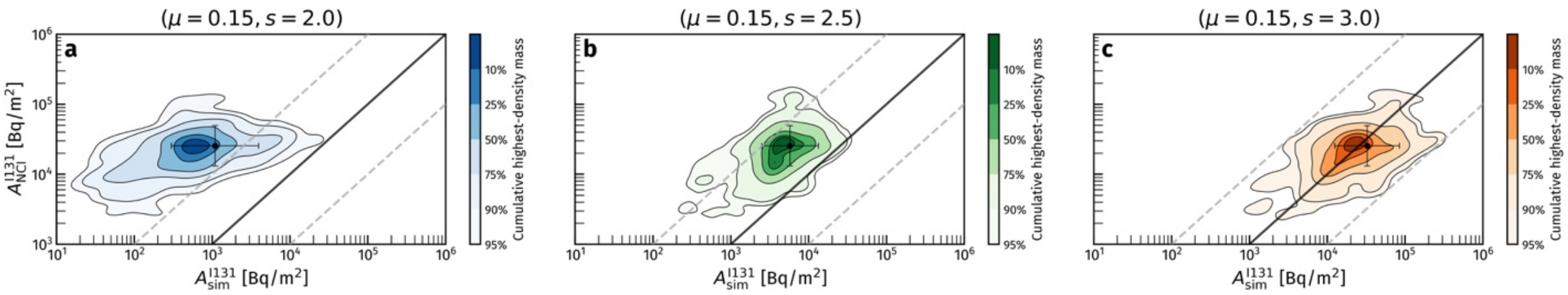

**Supplementary Figure 3. Comparison between HYSPLIT-reconstructed and National Cancer Institute (NCI) county-level iodine-131 deposition estimates for airburst tests, using three different log-normal particle size distributions.** Each panel shows the joint distribution of reconstructed (x-axis) versus NCI-reported (y-axis) iodine-131 deposition densities (in Bq $m^{-2}$), evaluated with a two-dimensional kernel density estimate (KDE) in log-space. Colored regions denote *highest-density regions* (HDRs) containing the densest 10%, 25%, 50%, 75%, 90%, and 95% of the probability mass. Results are shown for a fixed median particle diameter (μ = 0.15 μm) and three geometric standard deviations: s = 2.0 (**a**), s = 2.5 (**b**), and s = 3.0 (**c**). The solid black line indicates perfect 1:1 agreement, while dashed gray lines show ±1 order-of-magnitude deviations. Increasing geometric spread systematically improves agreement with NCI observational estimates, with the s = 3.0 distribution producing the closest match across counties. Points with error bars represent the mean reconstructed and observed deposition values with ±1σ log-space dispersion.

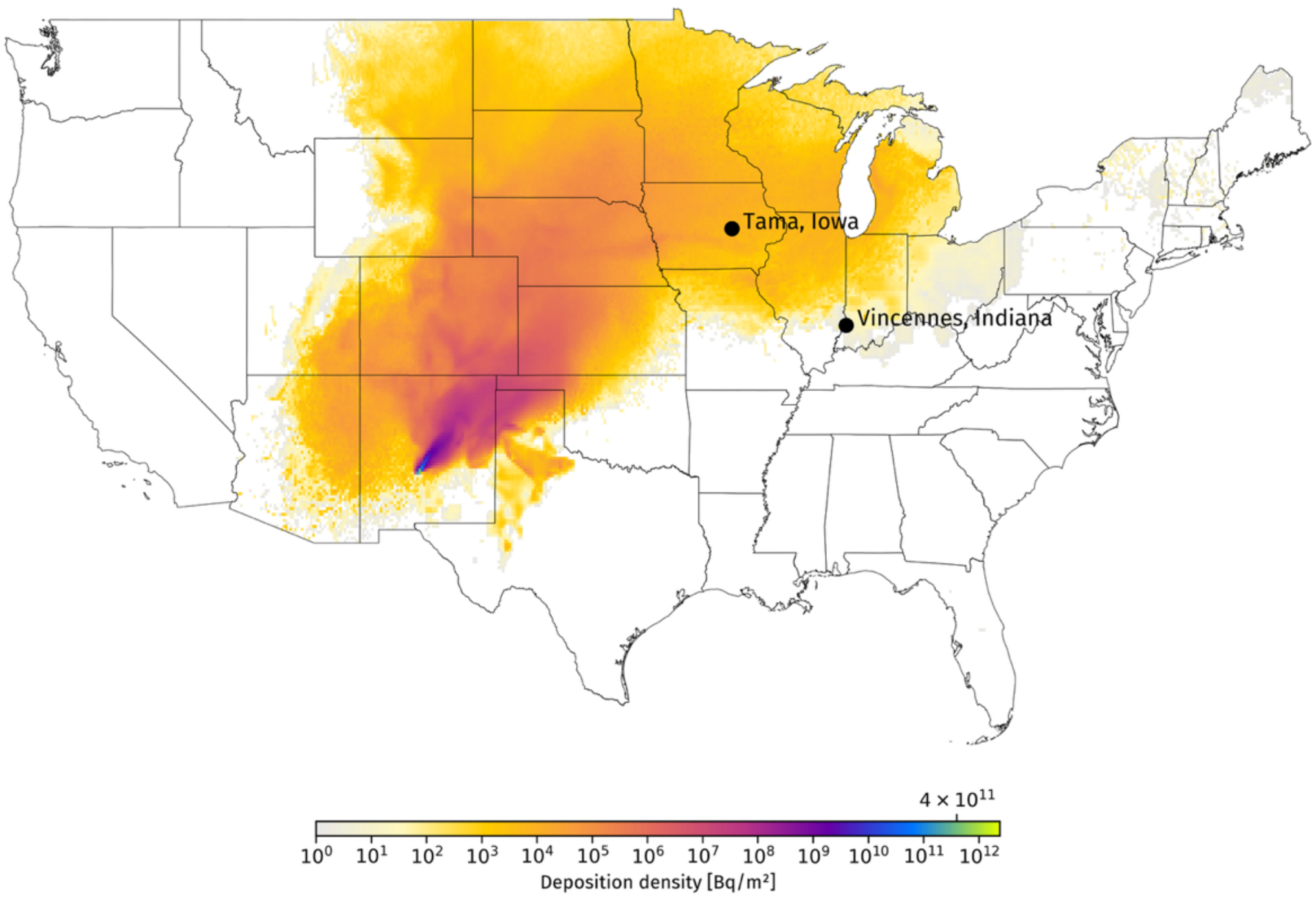

**Supplementary Figure 4. Reconstructed cumulative deposition density from the Trinity test six days after detonation, showing long-range fallout consistent with 1945 photographic film contamination.** In August 1945, the Eastman Kodak Company reported unexplained fogging on X-ray film packaged with strawboard manufactured at mills in Vincennes, Indiana, and Tama, Iowa. Radiochemical analysis of the contaminated board revealed beta-emitting isotopes with an approximately 30-day half-life, later attributed to fission products such as cerium-141. This was among the earliest documented instances of long-range fallout from Trinity. Our reconstruction of fission-product deposition (in Bq/m²) from the Trinity test shows northeastward transport of radioactive material, including deposition over Indiana and Iowa as of July 22, 1945 when fallout begins to reach Vincennes, Indiana.

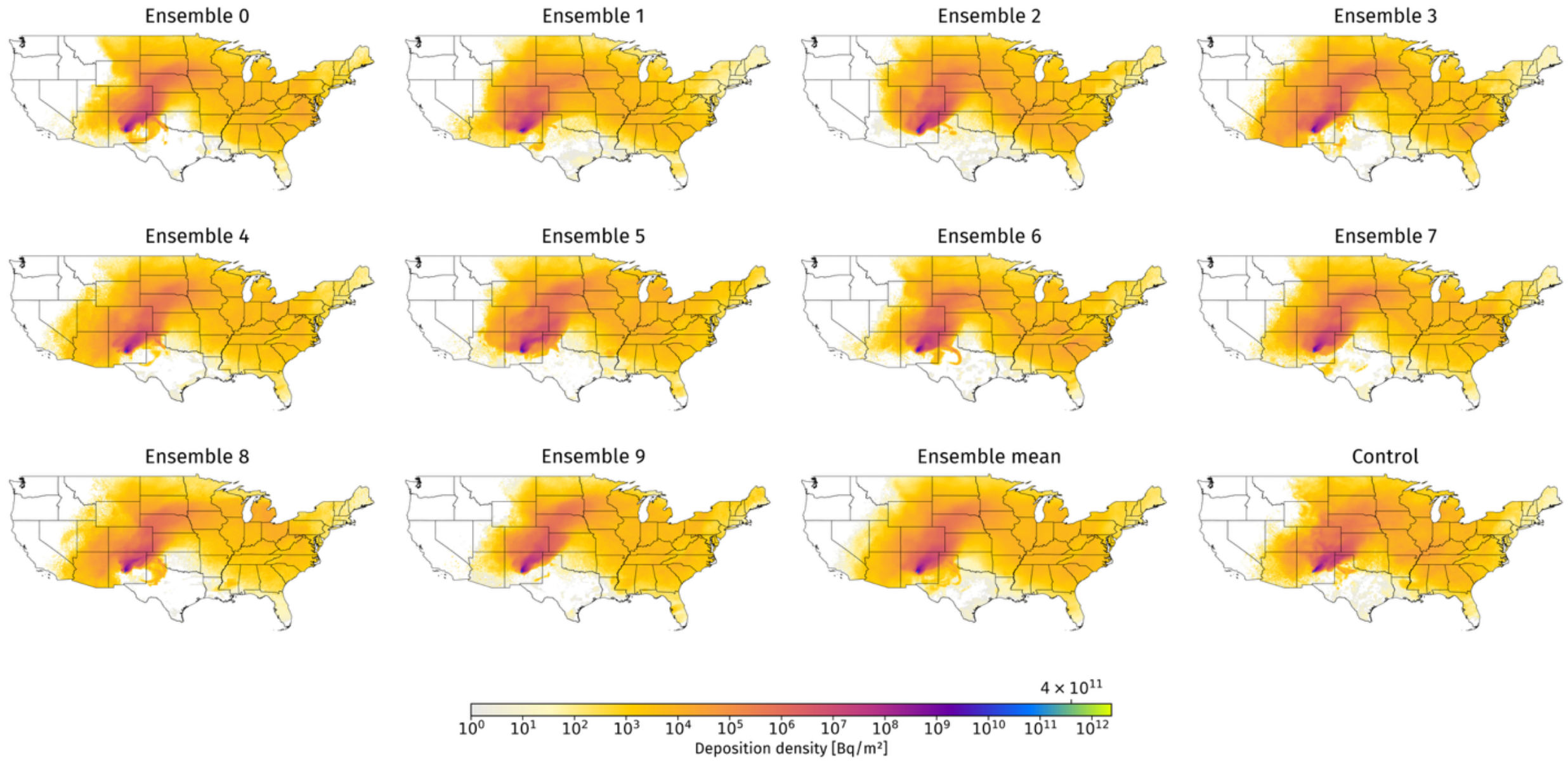

**Supplementary Figure 5. Ten-day cumulative fission-product deposition from the Trinity test using each ERA5 ensemble member, the ensemble mean, and the reanalysis (control) data set.** All reconstruction ensemble members use identical source terms, transport parameters, and spatial grids. While the spatial structure and intensity of fallout vary modestly between ensemble members, the overall geographic extent of deposition is consistent, with material reaching at least 45 of the United States in all members. These results highlight long-range transport features common to all realizations.

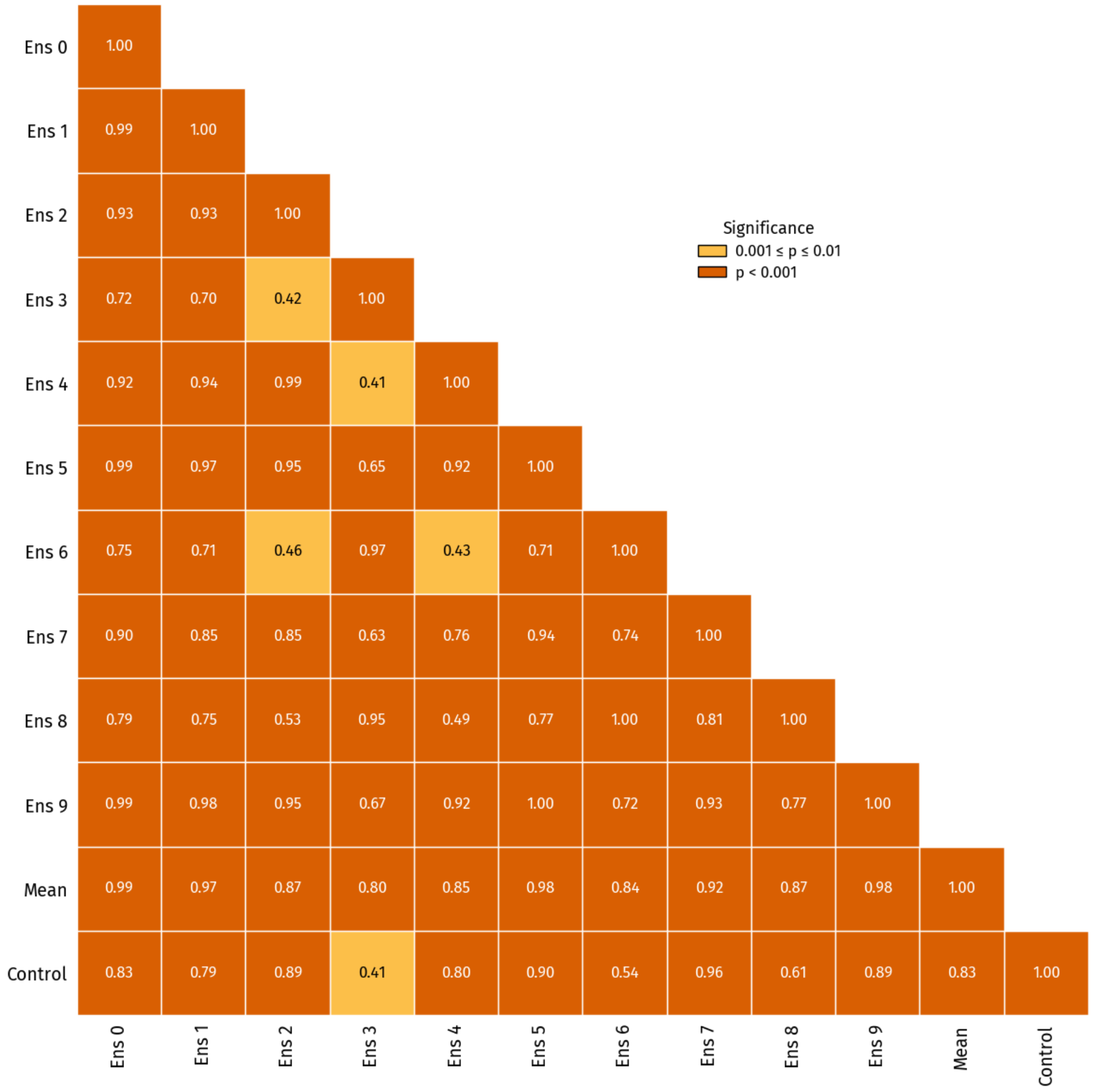

**Supplementary Figure 6. Pairwise Pearson correlation matrix for Trinity fallout deposition across ERA5 ensemble members and the control.** Matrix of Pearson correlation coefficients computed between the 10 ensemble members, the ensemble mean, and the control reconstruction for Trinity 10-day cumulative deposition fields. All values exceed the two-tailed significance threshold of 0.36 at $p = 0.01$, assuming 48 degrees of freedom (based on 5° × 5° independent blocks across the 20° × 60° reconstruction domain). Correlation values range from 0.41 to 0.99 between ensemble pairs, and from 0.41 to 0.96 between each ensemble member and the control, indicating strong spatial agreement in modeled deposition patterns across the ensemble spread.